%% file: elsarticle-main.tex
\documentclass[1p,12pt, sort&compress]{elsarticle}
\usepackage{amsmath,amssymb}
\usepackage{graphicx}% Include figure files
\usepackage{bm}% bold math
\usepackage{times}
\usepackage{xcolor}
\usepackage{lineno}
\usepackage{textcomp, gensymb}
\usepackage{hyperref}% add hypertext 
\usepackage{tabularx}
\usepackage{adjustbox}
\usepackage{subfigure}
\usepackage{siunitx}
\usepackage{float}

\newcommand{\Cf}{$^{252}$Cf}
\newcommand{\Mn}{$^{54}$Mn}
\newcommand{\Cs}{$^{137}$Cs}
\newcommand{\Co}{$^{60}$Co}
\newcommand{\Na}{$^{22}$Na}
\newcommand{\Li}{$^{6}$Li}
\newcommand{\Hy}{$^{1}$H}
\newcommand{\FoM}{\textsc{fom}}
\newcommand{\PMT}{\textsc{pmt}}
\newcommand{\IBD}{\textsc{ibd}}
\newcommand{\EJ}{\textsc{ej}{\footnotesize-299-50}}
\newcommand{\PSD}{\textsc{{psd}}}
\newcommand{\LO}{\textsc{{lo}}}
\newcommand{\PPO}{\textsc{{ppo}}}
\newcommand{\FWHM}{\textsc{fwhm}}
\newcommand{\pe}{\textsc{pe}}
\newcommand{\adc}{\textsc{adc}}
\DeclareSIQualifier\ee{ee}
\DeclareSIUnit\MeVee{\MeV\ee}

\newcommand{\LLNL}{Lawrence Livermore National Laboratory (LLNL) \renewcommand{\LLNL}{LLNL}}

\usepackage{verbatim}

\journal{Nuclear Instruments and Methods}

\begin{document}

\begin{frontmatter}



\title{Performance of large-scale $^6$Li-doped pulse-shape discriminating plastic scintillators}

\include{AuthorList2022}

% \author[inst2]{Author Two}
% \author[inst1,inst2]{Author Three}

% \affiliation[inst2]{organization={Department Two},%Department and Organization
%             addressline={Address Two}, 
%             city={City Two},
%             postcode={22222}, 
%             state={State Two},
%             country={Country Two}}

\begin{abstract}
A \Li-doped plastic scintillator with pulse-shape discrimination capabilities, commercially identified as \EJ{},  has been developed and produced at the kilogram-scale. A total of 44 \unit{\kg}-scale bars of dimensions \qtyproduct{5.5 x 5.5 x 50}{\cm} of this material have been characterized. Optical properties like light output and effective attenuation length have been found to be comparable to \Li{}-doped liquid scintillators. The scintillator \EJ{} shows good neutron detection capabilities with an effective efficiency for capture on \Li{} of approximately 85\%. Stability tests performed on two formulation variations showed no intrinsic degradation in the material or optical properties during several months of observations.
\end{abstract}

% \SI{14.6(3.0)}{\micro\second}


% \begin{highlights}
% \item First large-scale production of $^6$Li-doped PSD-capable plastic scintillators
% \item Plastics show comparable neutron detection capabilities, attenuation length and light output as liquid counterparts
% \item No significant degradation of the plastics has been found during the study.
% \end{highlights}


\end{frontmatter}


\input{Introduction}

\input{ScintillatorDescription}
\input{CharacterizationSetup}
\input{ScintillatorPerformance}
\input{Conclusions}

\bibliographystyle{elsarticle-num} 
\bibliography{references}





% \end{thebibliography}
\end{document}

%% file: AuthorList2022.tex
% !TEX root = main.tex
% \address{, Livermore, CA, USA} 

% \author{C.~Roca$^{7}$,
% N.~S.~Bowden$^{7}$,
% S.~A.~Dazeley$^{7}$,
% M.~J.~Ford$^{7}$,
% V.~Li$^{7}$,
% F.~Sutanto$^{7}$,
% N.~P.~Zaitseva$^{7}$,
% Mendenhall,
% Durham?,
% Eljen authors.
% }

\author[LLNL]{C.~Roca\corref{cor1}}
\cortext[cor1]{Corresponding author}
\ead{rocacatala1@llnl.gov}

\author[LLNL]{N.~S.~Bowden}
\author[LLNL]{L.~Carman}
\author[LLNL]{S.~A.~Dazeley}
\author[LLNL]{S.~R.~Durham}
\author[Eljen]{O.~M.~Falana}
\author[LLNL]{M.~J.~Ford}
\author[LLNL]{A.~M.~Glenn}
\author[Eljen]{C.~Hurlbut}
\author[LLNL]{V.~A.~Li}
\author[LLNL]{M.~P.~Mendenhall}
\author[Eljen]{K.~Shipp}
\author[LLNL]{F.~Sutanto}
\author[LLNL]{N.~P.~Zaitseva}

\affiliation[LLNL]{organization={Lawrence Livermore National Laboratory},
            city={Livermore},
            state={CA},
            postcode={94550}, 
            country={USA}}

\affiliation[Eljen]{organization={Eljen Technology},
            city={Sweetwater},
            state={TX},
            postcode={79556}, 
            country={USA}}

%% file: Introduction.tex
% Introduction
\section{Introduction}
Liquid scintillators have been the standard technology for reactor-antineutrino detection for decades~\cite{Reines:1953pu,Declais:1994su,CHOOZ:1997cow,Boehm:1999gk,KamLAND:2002uet,DoubleChooz:2011ymz,DayaBay:2012fng,RENO:2012mkc,PROSPECT:2018dtt}. Besides enabling antineutrino interaction through Inverse Beta Decay (\IBD{}), many liquid scintillators can be used to discriminate between particle ionization density, e.g. electronic vs nuclear recoil, via Pulse-Shape Discrimination (\PSD{}) \cite{Birks:1964}. Further particle identification can be provided for thermal-neutrons by the addition of dopants that have specific neutron capture reactions. For example, inclusion of \Li{} in a detector system can provide a distinct capture tag via a combination of energy and \PSD{} selections~\cite{Declais:1994su,KIFF2011412,SoLid:2018jas,PROSPECT:2018dtt,Haghighat:2020,NuLat:2019,Sutanto:2021}. 

While liquid scintillators are relatively inexpensive, easy to manufacture,  and have demonstrated high performance, they require consideration of potential safety hazards, handling, transportation, and maintenance of an appropriate operating environment. Safety, regulatory, practicality, and mobility requirements can put constraints on the development of novel detectors, particularly the physical state of the detection medium. Plastic scintillators alleviate some of these concerns.

It is only recently that such plastics have been developed with \PSD{} capability and \Li{}-doping~\cite{Zaitseva:2012, Zaitseva:2013, Breukers:2013, Cherepy:2015, Zaitseva:2018, Frangville:2019}. A key challenge in their production is achieving a balance between the solubility of the components with the main monomer and the careful control of curing conditions, all while maintaining satisfactory light output and optical properties. 
Introducing \Li{} as a thermal-neutron capture agent to \PSD{} plastics, e.g. to achieve similar capabilities to doped liquid scintillators~\cite{PROSPECT:2020sxr}, further complicates the formulation and production process. Unlike common primary dyes like meta-terphenyl (mTP) or 2,5-diphenyloxazole (\PPO{}), most organic salts containing \Li{} as the cation do not have solubility in monomers like styrene, commonly used as a base for plastic scintillator production. Recent efforts have identified \Li{}-bearing compounds, scintillator formulations, and curing techniques that can produce kilogram-scale \Li{}-doped \PSD{}-capable plastics with reasonable performance~\cite{Ford:2023, Sutanto:2021}. Eljen Technology has made further significant advances in performance and scalability for this material, which has been commercially labelled as \EJ{}. 

The demonstration of adequate optical performance and neutron identification capabilities by \EJ{} could make it an appropriate candidate for future safeguards and reactor monitoring experiments using antineutrino detection that require near-field, readily deployable systems~\cite{Carr:2018, Bernstein:2020}. The ability to detect fast-neutrons and distinguish them from thermal-neutrons and gamma-rays also makes \EJ{} an ideal candidate for neutron spectroscopy and multiplicity counting applications. 

In this work, we characterize the scintillation and neutron detection performance of \EJ{}. We primarily focus on long bar geometries suitable for segmented reactor-antineutrino detectors. Sec.~\ref{Sec:AttLength} describes the process to characterize the effective light output and attenuation length of the plastic by means of using a collimated gamma-ray source (\Na{}). In Sec.~\ref{Sec:NeutronStudies}, the response of the plastic when exposed to a mixed neutron/gamma-ray source (\Cf{}) is detailed. Quantities such as capture time, \PSD{} figure of merit, and thermal-neutron capture fraction are measured and compared with simulations. Finally, in Sec.~\ref{Sec:Stability}, we discuss the results of aging tests and the long-term stability of the plastic.

%% file: ScintillatorDescription.tex
\section{Scintillator Description}
Plastic scintillators have gained attention as candidate materials for reactor-antineutrino detection due to the  progress made in developing stable samples with \PSD{} capabilities \cite{Zaitseva:2012, Zaitseva:2018, NuLat:2019, Sutanto:2021}. 
%The decade-long expertise of a dedicated materials group at Lawrence Livermore National Laboratory has allowed for the production of plastics with different compositions for testing.
Desired properties during the production of such plastics are long-term stability, long attenuation lengths to allow for large-scale detection elements, and relatively simple processing requirements. At the same time, maintaining good \PSD{} and thermal-neutron capture capabilities is crucial for antineutrino detection. 

%Plastic scintillators have gained popularity recently as candidates for neutrino detection due to the significant progress made in synthesizing stable samples with \PSD{} capabilities \cite{Zaitseva:2012, Zaitseva:2018}. The decade-long expertise of a dedicated materials group at Lawrence Livermore National Laboratory has allowed for the production of plastics with different compositions for testing. Desired properties during the production these plastics are long-term stability, long attenuation lengths to allow for large-scale production  and relative simple processing requirements. At the same time, keeping \PSD{} and thermal neutron capture capabilities is crucial for neutrino detection.

With these considerations in mind, \Li{}-doped \PSD{} plastic scintillator formulations and casting processes for producing kilogram-scale detection elements have been  developed and tested at LLNL \cite{Ford:2023}. Eljen Technology have further advanced this class of material, in the form of \EJ{}, and can now produce elements at the meter-scale with mass greater than 4~kg. This material contains about $0.1\%$ \Li{} by weight and is based on polyvinyl toluene, containing \PPO{} as a primary dye. \Li{} is introduced in the form of an aliphatic salt, whose solubility is enhanced by use of a comonomer.

\begin{figure}[H]
\centerline{\includegraphics[width=0.7\textwidth]{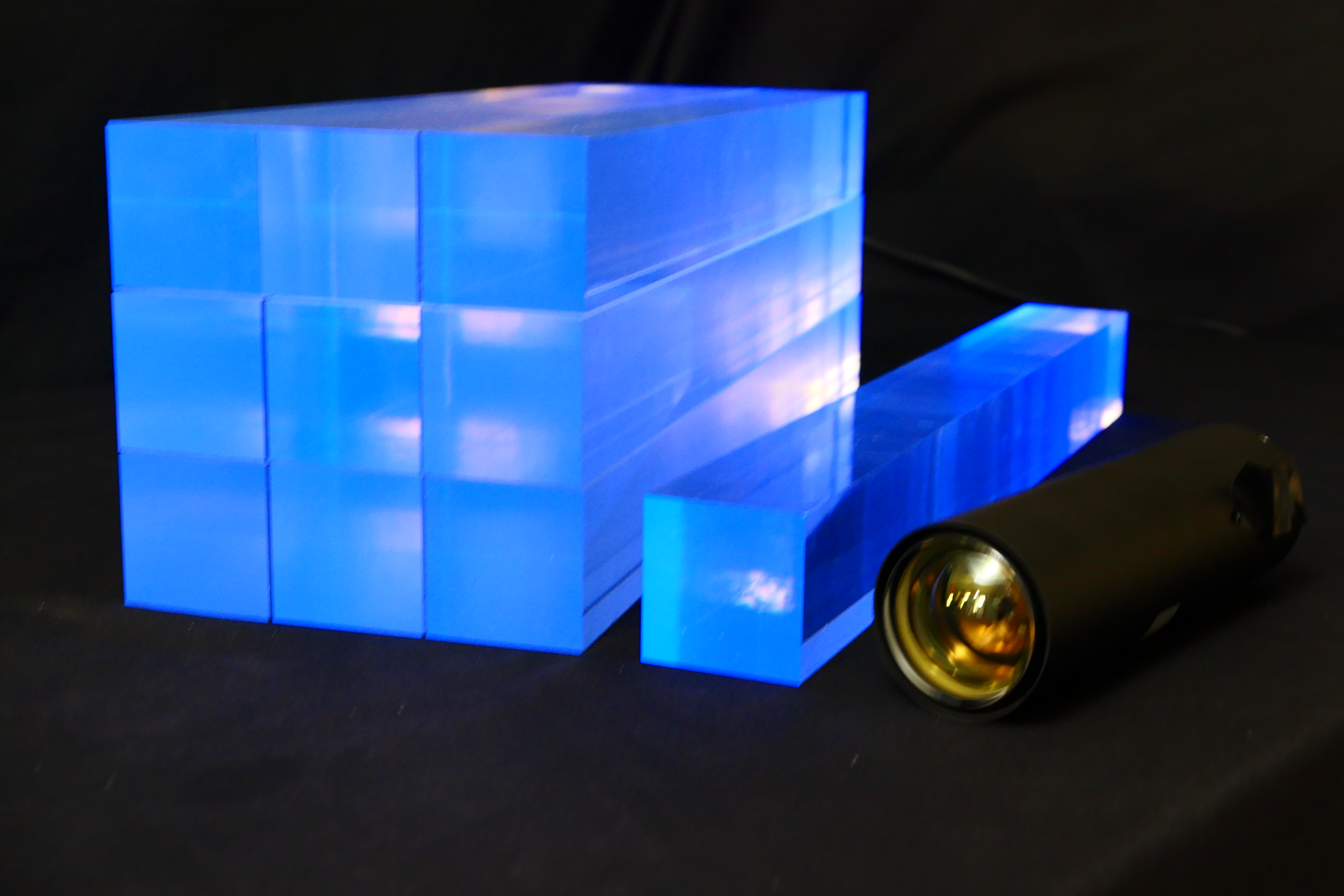}}
\caption{A subset of 10 \EJ{} bars of \qtyproduct{5.5 x 5.5 x 50}{\cm} dimension produced by Eljen Technology. Next to them, for scale, a Hamamatsu 2-inch-diameter H11284-100 photomultiplier tube assembly. Picture taken under UV-light illumination.}
\label{Fig:PSDBars}
\end{figure}

In this work, we report on the performance of 44~bars of \EJ{} with dimensions of \qtyproduct{5.5 x 5.5 x 50}{\cm} from a development production run as well as a set of 2-inch-diameter cylindrical samples. All these were produced by Eljen Technology between November 2021 and September 2022, during which period minor variations in formulation and process were implemented to improve the material properties. A photograph of several of the \SI{50}{\cm} long bars is shown in Fig~\ref{Fig:PSDBars}. After the characterization effort described here, 36 out of 44 of these \SI{50}{\cm} bars have been installed as active material of the Reactor Operations Antineutrino Detection Surface Testbed Rover (ROADSTR) detector~\cite{ROADSTR:2021}.

% \subsection{Kilogram-scale production by Eljen Technology}

% Building from previous work at \LLNL{}, Eljen Technology (ET) have developed LiPS formulations with improved performance, producing detection elements as large as 1~meter. In this work we report on the performance of 44~bars with dimensions of $5.5\times5.5\times50$~cm from a development production run, as well as a bar of dimensions $6.0\times6.0\times100$~cm. ET's formulation is designated as \EJ{} and minor variations in formulation and process occurred throughout the production run to improve the materials properties. A photograph of several of the 50~cm long bars is shown in in Fig~\ref{Fig:PSDBars}. %The results of such effort have are summarized in this publication. 
% After the characterization effort described here, 36 of these \SI{50}{\centi\metre} bars have been installed as active material of the ROADSTR detector~\cite{ROADSTR:2021} during its current commissioning phase.

%% file: CharacterizationSetup.tex
\section{Experimental Description} \label{Sec:ExperimentalDescription}

Properties of the \EJ{} \PSD{} plastic are initially tested using small 2-inch-diameter cylindrical samples wrapped in a diffuse Teflon reflector and optically coupled to a 2-inch-diameter \textsc{r\footnotesize6231-100-\normalsize{sel}} \PMT{}. The light output and \PSD{} performance of the material were measured using \Cs{} and \Cf{} sources respectively. This setup will be referred as the \textit{small} configuration.

The bar castings are tested in batches of four, by means of a support system that holds them parallel in a \numproduct{2 x 2} grid configuration. Each bar is read out by a pair of Hamamatsu 2-inch \textsc{h\footnotesize{11284-100}} bi-alkali \PMT{} assemblies \cite{Misc:PMT}. Contact between scintillator and \PMT{} is maintained by inward pressure from a series of contracted springs attached to the support frame.

Bars are characterized in several configurations. In the \textit{wrapped} configuration, the bars are covered with 3M\texttrademark ~\textsc{df\footnotesize{2000}\normalsize{ma}} specular reflector \cite{Misc:Wrap}, leaving a small ($\lesssim$ \SI{0.1}{\mm}) air gap between the bar and the reflector to allow for efficient optical transport of the fraction of scintillation photons that meet the conditions for Total Internal Reflection (\textsc{tir}) at the bar-air interface. \PMT{} assemblies are grease-coupled to bars via a layer of {\textsc{ej}\footnotesize-550} optical grease  \cite{Misc:EJ550}. This configuration provides a high light-collection efficiency and, in general terms, corresponds to the optimal setup for any 2D detector. This configuration has been used to measure the \textit{effective} light output of the bars, as described in Sec.~\ref{Sec:EnergyScale}, as well as to obtain the \PSD{} measurement presented in Sec.~\ref{Sec:NeutronStudies}.

The possibility of bar-to-bar inconsistency in the wrapping and grease-coupling process, e.g. flatness of specular reflector strips or grease coupling uniformity, had the potential to affect the comparison of bar properties. Thus, for the attenuation length measurements described in Sec.~\ref{Sec:AttLength}, the plastic bars are left in the \textit{bare} configuration, i.e. measured without coupling grease or specular reflector wrap. Although this choice results in lower light collection, it allows for a more reliable bar-to-bar comparison. Additionally, neutron-capture efficiency studies described in Sec.~\ref{Sec:NeutronStudies}, benefit from a simple setup that can be easily replicated in simulation. For this reason, and given that their addition does not bring any particular benefit to the efficiency measurements, these also used the \textit{bare} configuration.

Most of the bar-characterization measurements described here use gamma-ray sources that are housed in a \SI{5}{\cm}-wide lead source holder that collimates gamma emissions through up to four \SI{0.5}{\cm}-diameter circular apertures to provide spatially localized energy depositions. To characterize optical collection as a function of interaction position, a linear-translation stage is configured to transport the source holder longitudinally along the main axis of the bar. Both the stage and the data acquisition are controlled remotely through scripted routines, providing automated data-collection at specific source positions.

%\todo{Here we should descirbe all configurations and their motivation, e.g. reflector and coupling with grease to get light yield, PSD, and attenutaon relevant for experitments; naked bars and no coupling for comparison between batches of bars}

\subsection{Data acquisition}

In the \textit{small} configuration, \PMT{} signal is recorded using a 14-bit resolution CompuScope 14200 \cite{Misc:CompuScope} waveform digitizer with a sampling rate of \SI{200}{\mega\hertz}. Readout of the \PMT{}s during the bar measurements is performed using a CAEN V1725S digitizer with \SI{250}{\mega\hertz} sampling frequency and 14-bit resolution \cite{Misc:CAEN}. 

Waveform shape depends on the type of event being recorded. Heavily ionizing particles, like neutron-induced recoil-protons or the products from neutron-captures in \Li{} (\Li{}$(n,t)\alpha$), generate a population of long-lived triplet states in the scintillator, leading to slower light emission and longer pulses than singlet states, induced by electron-like ionization from gamma-ray events. The \PSD{} parameter quantifies this effect:
\begin{equation} \label{Eq:PSD}
    \PSD{} = Q_{\text{tail}}/Q_{\text{tot}},
\end{equation}
where $Q_{\text{tot}}$ is the total integrated charge in a given waveform and $Q_{\text{tail}}$ is the integrated charge in a specific window corresponding to the \textit{tail} of the pulse. The time-integration windows for $Q_{\text{tot}}$ and $Q_{\text{tail}}$ are [-8, 400]~ns and [36, 400]~ns with respect to the half-height of the waveform's leading edge, respectively. Using these definitions, the \PSD{} parameter is calculated for each event using Eq.~\ref{Eq:PSD}. 

These time-integration windows were chosen to maximize the separation Figure of Merit (\FoM{}) between nuclear and electronic-recoil, which can be interpreted as an indication of the gamma-neutron event discrimination capabilities of the plastics. It is commonly defined as:

\begin{equation} \label{Eq:FoM}
    \FoM{} = \frac{\mu_{\gamma}-\mu_n}{\FWHM{}_{\gamma}+\FWHM{}_{n}} =  0.425\cdot\frac{\mu_{\gamma}-\mu_n}{\sigma_{\gamma}+\sigma_{n}}
\end{equation}
where $\FWHM{}_i = 2.35\sigma_i$ corresponds to the Full Width Half Maximum of the band, approximated as a Gaussian distribution.

The high voltage bias applied to each \PMT{} is chosen to result in matched gain. Gain curves were obtained from LED pulser calibrations that determined the analog-digital conversion units (\adc{}) response for single photo-electron (\pe{}) incidences as a function of applied bias. Further, the chosen gain setting (\SI{40}{\adc/\pe}) results in a linear response throughout the \qtyrange{1}{8}{\MeV} energy range. A trigger acquisition threshold of \SI{120}{\adc} (\SI{3}{\pe}) is also applied.

\begin{figure}[h]
    \centering
        {\includegraphics[width=0.9\textwidth]{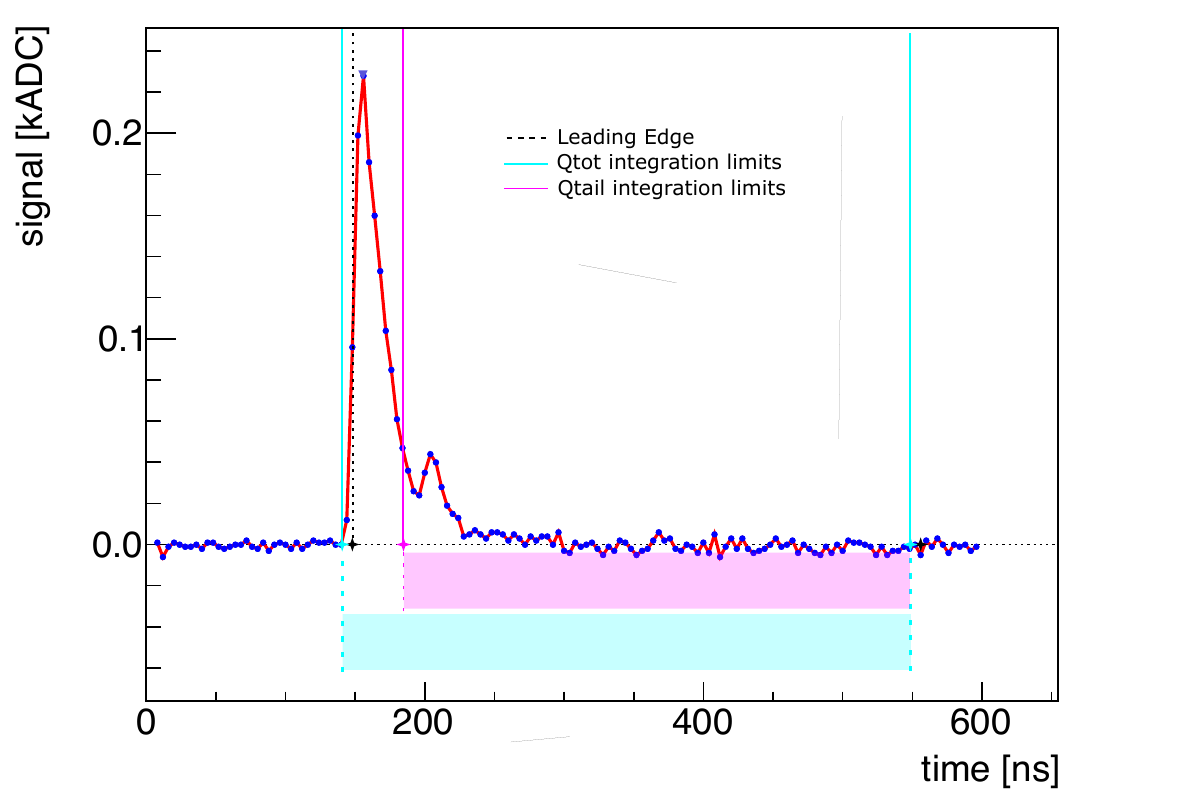}}
        \caption{Example of a \PMT{} waveform sampled in \SI{4}{\nano\second} intervals. Cyan lines delimit the $Q_{tot}$ integration boundaries while magenta specifies the boundary between rise and tail of the waveform. In black, crossed by a dashed line, the waveform's leading edge is depicted.}
        \label{Fig:PSDWaveform}
\end{figure}

%% file: ScintillatorPerformance.tex
\section{Scintillator Performance}
In this section, we describe the measurement of the main performance metrics of the \EJ{} scintillator.

\subsection{Small Sample Performance Measurement} \label{Sec:Small}
\begin{figure}[H]
    \centering
        \includegraphics[width=0.74\textwidth]{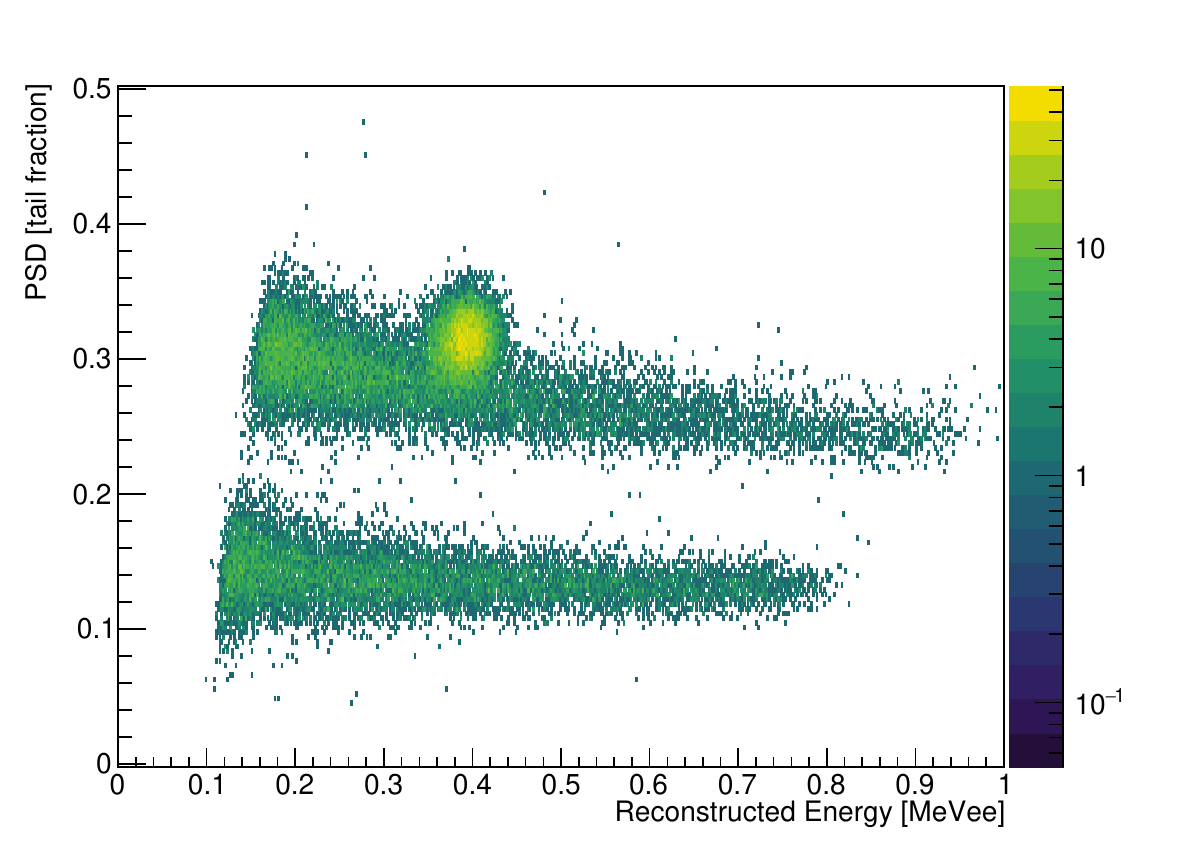}
        \caption{\PSD{} distribution in terms of reconstructed energy for \Cf{} data on the 2-inch-diameter cylindrical elements of \EJ{}.}
        \label{Fig:MoneyPlot2inch}
\end{figure}

As described in the previous section, the first set of measurements of \EJ{} are done for a single sample in the \textit{small} configuration. This configuration allows us to compare \EJ{} with other benchmark plastic and liquid scintillator measurements, also performed in small volumes. The distribution of reconstructed energy versus \PSD{} is presented in Fig.~\ref{Fig:MoneyPlot2inch}. Good band separation is observed for fast-neutron and gamma-like events,  as is for the neutron-capture feature at \SI{\sim 0.40}{\MeVee}. The \PSD{} \FoM{} (Eq.~\ref{Eq:FoM}) is 2.11 between \qtylist{0.35;0.50}{\MeVee}. This result compares favorably with previous unloaded \PSD{} plastic formulations like {\textsc{ej-}\footnotesize{299-33}} \cite{Lawrence:2013}, with $\FoM{}\sim0.9$ for fast-neutron and gamma bands in the \SIrange{0.35}{0.5}{\MeV} energy region. \EJ{} showcases a more comparable \FoM{} to recent unloaded formulations like {\textsc{ej-}\footnotesize{276}}\cite{Zaitseva:2018}. When comparing plastics to liquids of the same size, like {\textsc{ej-}\footnotesize{309}} in a 2-inch diameter cylindrical cuvette, a larger discrepancy is observed in favor of the liquid scintillator, which \FoM{} ranges between \numlist{3.18;3.95} for two different \Li{}-doped variants \cite{Zaitseva:2023}. Such difference comes mainly from broader \PSD{} distributions for plastics due to less homogeneous energy transfers in the material \cite{Zaitseva:2018}.

\subsection{Light Collection and Attenuation Length in Long-Bars} \label{Sec:AttLength}
As shown in the previous section, \EJ{} shows good performance when cast as small samples. However, applications like antineutrino detection or neutron multiplicity counting require larger mass and/or surface area. In such geometries, light collection may not be spatially uniform. Longer light propagation paths result in increasing losses through absorption or scattering in the material and at optical interfaces. Additionally, variation in travelled path length for different emission angles may cause timing dispersion that affects the measured scintillation time profile. All of these can degrade \PSD{} performance, making it important to test and characterize the properties of \EJ{} in the form of long-bar castings for large detector applications. 

In this work, we estimate an \textit{effective} attenuation length ($\lambda$) and an \textit{effective} light output ($\ell$) for \EJ{}. These estimations are not a representation of the intrinsic properties of the scintillator but rather a combined effect of optical properties with light-loss mechanisms. To do these calculations, a set of measurements with a \Na{} source are performed. The source is placed inside a lead collimator on top of a linear stage, as described previously in Sec.~\ref{Sec:ExperimentalDescription}. The stage is configured to take 20 equidistant measurements across the longitudinal axis of the bar, providing a position-dependent response. To a reasonable approximation, the collected signal $S_0 (S_1)$ for a given energy deposition $E_{\text{dep}}$, e.g. the Compton-edge, decreases (increases) exponentially as the source moves away from (closer to) the readout $\PMT{}_0$ ($\PMT{}_1$) on the bar ends:
\begin{align}
    S_{0} &= \kappa \cdot \exp{(-x/\lambda)} \label{Eq:S0} \\
    S_{1} &= \kappa \cdot \exp{((x-L)/\lambda)} \label{Eq:S1},
\end{align} 
where $x$ is the position of the source in the longitudinal axis relative to the end of the bar, and $L = \SI{50}{\cm}$ is the length of the bar. The response scale $\kappa$, measured in \adc{}, is proportional to the \textit{effective} light output $\ell$:
\begin{equation} \label{Eq:k}
    \kappa = \ell \cdot G \cdot E_{\text{dep}},
\end{equation}
and depends on the \PMT{} gain $G$, measured in \unit{\adc\per\pe}. If both \PMT{}s are gain-matched, their respective $\kappa$ factors are equivalent. Note that, by definition, $\ell$ in Eq.~\ref{Eq:k} already includes all unaccounted detection inefficiencies, and thus is measured in \textit{effective} \unit{\pe\per\MeV}. Both $\ell$ and $\lambda$ can then be extracted from Eqs.~\ref{Eq:S0} and \ref{Eq:S1} through the analysis of the multi-position \Na{} measurements:
\begin{enumerate}
    \item $S_0$ and $S_1$ can be combined in the following way:
\begin{equation} \label{Eq:lnS0S1}
    \ln{\frac{S_1}{S_0}} = \frac{2x}{\lambda} - \frac{L}{\lambda}.
\end{equation}
In Eq.~\ref{Eq:lnS0S1}, the signal ratio $\ln{S_1/S_0}$ evolves linearly with the source position $x$ and it is distributed between $\ln{S_1/S_0} = \pm L/\lambda$ for $x = 0,L$ respectively. This variable provides a compact observable that can be used to calculate $\lambda$. 
\item Similarly, the geometric mean of the two \PMT{} signals can be computed as follows:
\begin{equation} \label{Eq:sqrtS0S1}
    \sqrt{S_0 \cdot S_1} = \kappa \cdot \exp{(-L/2\lambda)},
\end{equation}
where the light collection is position-independent, at least in the regime where Eqs.~\ref{Eq:S0} and \ref{Eq:S1} are valid approximations. From Eq.~\ref{Eq:sqrtS0S1} it can also be noted that light collection degrades as the $L/\lambda$ ratio increases, showcasing the higher attenuation rate for longer photon paths. Once $\lambda$ is known, we can extract $\kappa$ by substitution using Eq.~\ref{Eq:sqrtS0S1} for every position. The \textit{effective} light output $\ell$ is then calculated by proxy using Eq.~\ref{Eq:k}. It is useful to scale the segment signal by $E_\text{dep}$ for the Compton-edge as follows:
\begin{equation} \label{Eq:scale}
    \varepsilon = \sqrt{S_0 \cdot S_1}/E_\text{dep} = \ell \cdot G \cdot \exp{(-L/2\lambda)},
\end{equation}
where the resulting variable $\varepsilon$ can be understood as the energy scale of the system.
\end{enumerate}

\begin{figure}
    \centering
     \includegraphics[width=0.75\textwidth]{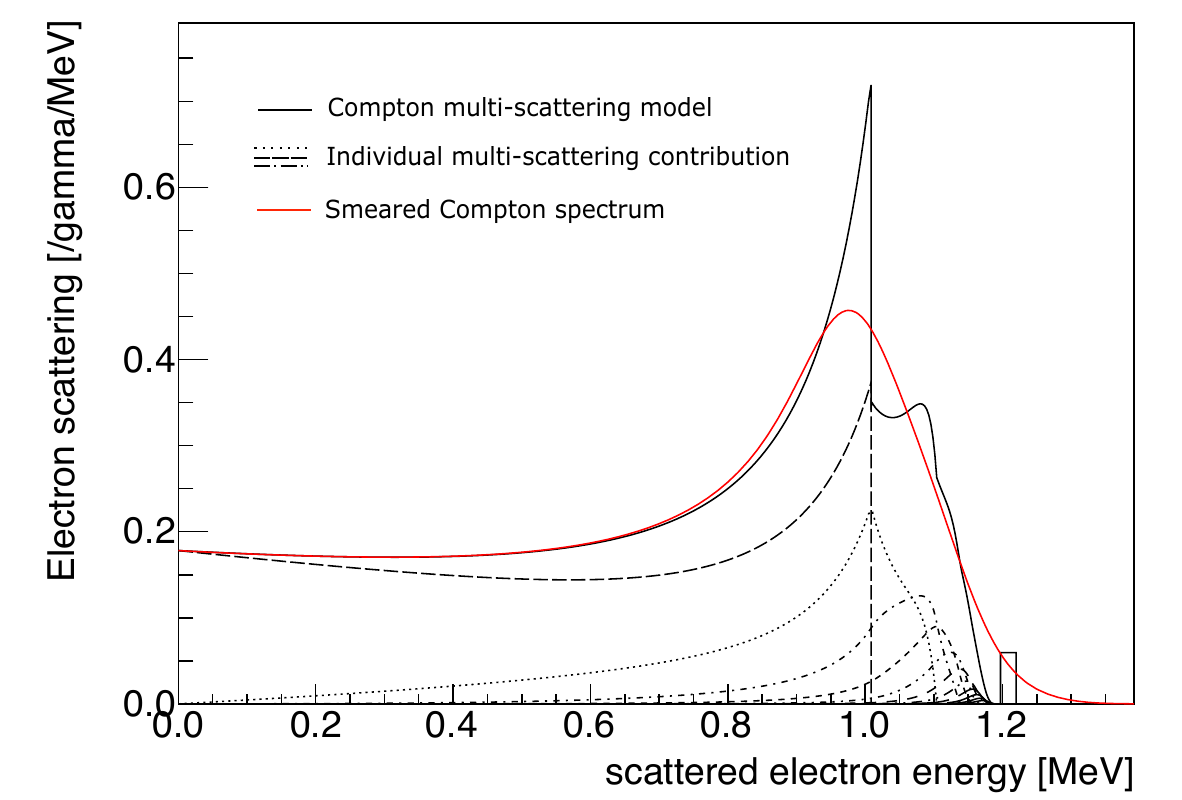}
    \caption{\Na{} gamma-ray electron-equivalent energy spectrum. Theoretical Compton multi-scattering distribution showed in solid black, with individual scattering contributions showed in dashed black. Solid red represents the folded detected response, defined by theoretical distribution with Gaussian smearing. The single bin at \SI{1.22}{\MeV} represents the residual distribution of full-energy deposition events through photoelectric effect. }
    \label{Fig:comptonEdge}
\end{figure}

However, before analysing the data to extract these properties, there are two pre-analysis tasks that need to be performed. Firstly, a Compton-edge fit is designed. Full-energy gamma-ray depositions rarely happen in the relatively small volume of single bars used in the present study. Furthermore, organic scintillators are mainly composed by organic chains of C and H (low-Z elements), that greatly reduce the cross-section for photoelectric effect, in favour of gamma-electron scatterings. For this reason, the most common occurrence is that the emitted gamma-ray scatters once in the material before leaving the active volume. This is represented in the scattered-electron energy distribution by a continuum with a sharp cut-off. This cut-off is often referred as Compton-edge \cite{Knoll:2000}. However, there is still a non-negligible probability that multiple scatterings are observed in one of the bars. Such Compton multi-scattering distribution can be observed in Fig.~\ref{Fig:comptonEdge} as a solid black line. This distribution has been obtained by numerically integrating the multi-scattering and photoelectric cross-sections. The individual non-solid lines represent the decreasing contributions to the total cross-section for an increasing number of potential scatterings before the gamma-ray either escapes the volume or is completely absorbed by the photoelectric effect. If we apply a detection smearing to account for light collection within the bar's geometry, we obtain the red curve. This smearing depends on four experimental parametrizations: 
\begin{itemize}
    \item Effective material thickness for gamma-ray scattering, representing the average number of scatterings per event.
    \item Event rate, working as a normalization factor.
    \item Energy resolution, a measure of the Compton-edge broadening.
    \item Energy scale, defined in Eq.~\ref{Eq:scale}. This parameter encapsulates information about the Compton-edge position.
\end{itemize}

\begin{figure}[H]
    \centering
    \subfigure[Position cut]{\includegraphics[width=0.49\textwidth]{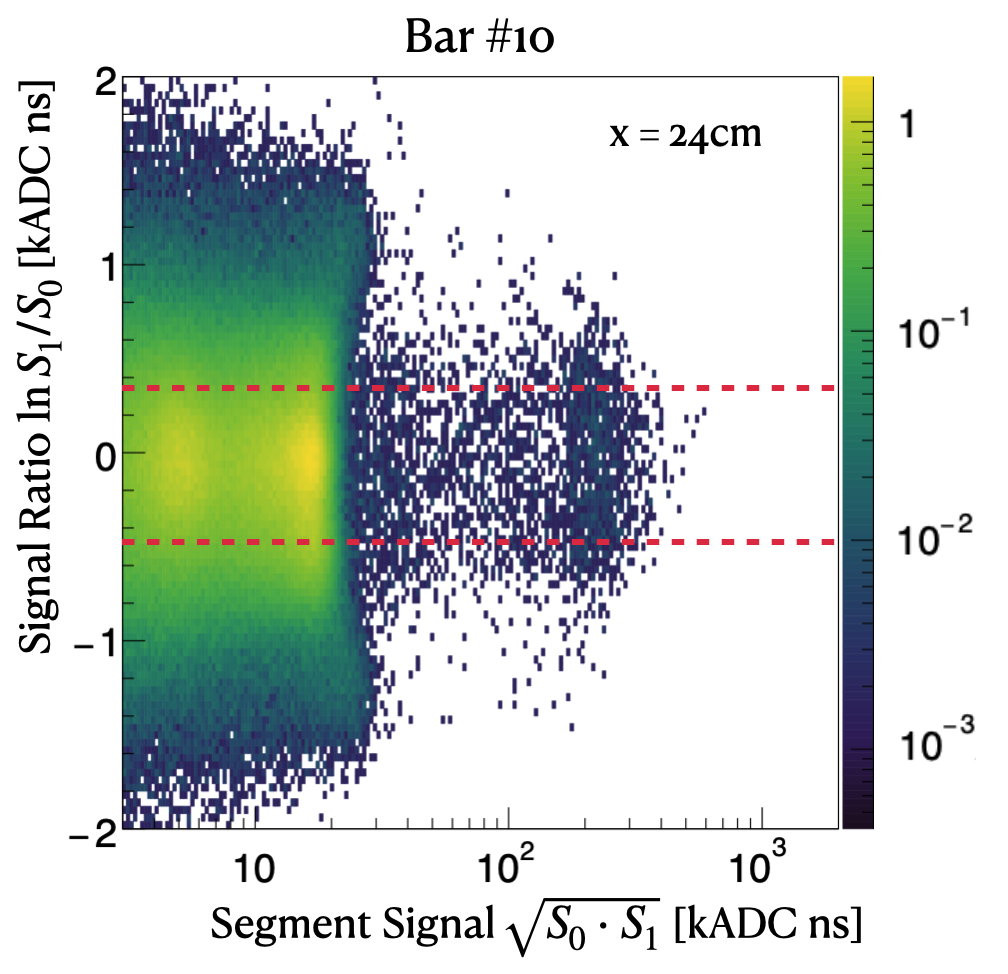} \label{Fig:positionCut}} 
    \subfigure[Measured Compton-edge]{\includegraphics[width=0.44\textwidth]{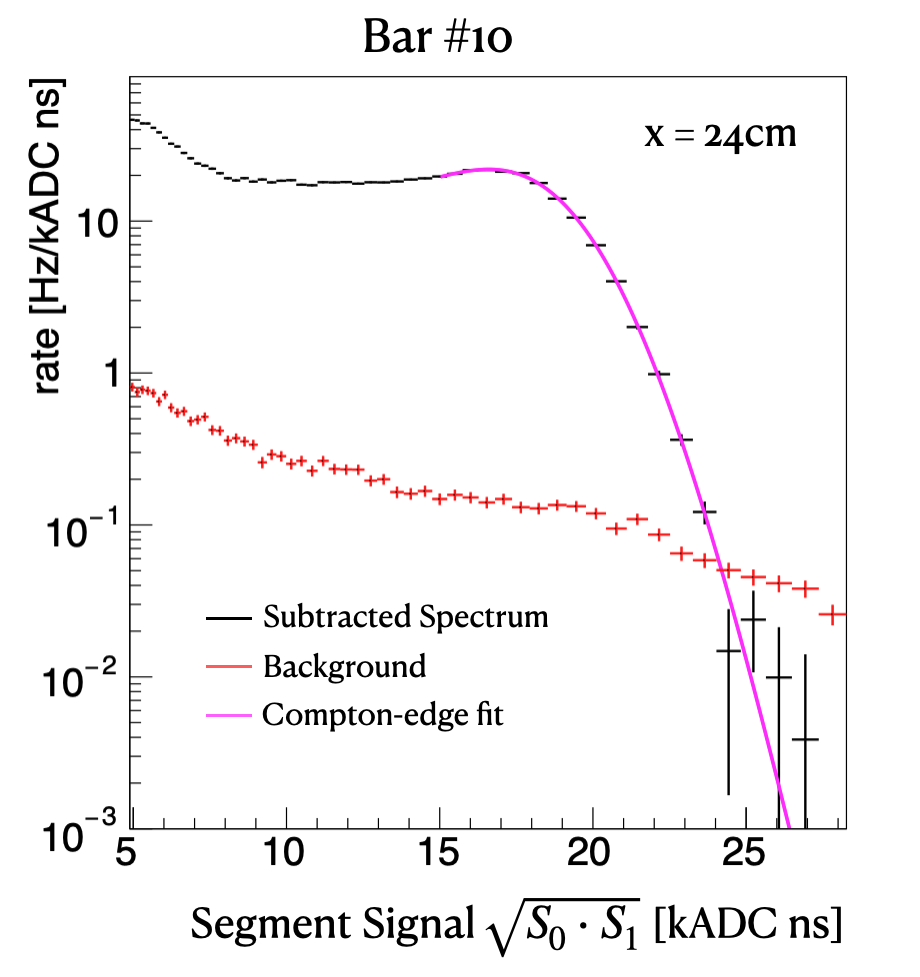} \label{Fig:comptonEdgeData}} 
    \caption{(a) Relationship between signal ratio $\ln{S_1/S_0}$ and geometric mean $\sqrt{S_0 \cdot S_1}$ for \Na{} data on bar \#10 at $x = \SI{24}{\cm}$. This position roughly corresponds to the middle point between both \PMT{}s and thus $\ln{S_1/S_0} \sim 0$ for properly collimated gamma-rays. Dashed lines correspond to the upper and lower cuts applied to the distribution, one standard deviation around the main peak in the $\ln{S_1/S_0}$ axis (b) Measured spectrum for \Na{} using bar \#10, after applying a position cut. Background (red) has been subtracted from data. The Compton-edge fit is displayed in magenta.}
\end{figure}

Secondly, a position cut for \Na{} data is defined. During data taking, the lead collimator is positioned in such a way that the beam of gamma-rays emitted by the source hits the bars perpendicularly, i.e. at the same position $x$ as the source across the longitudinal axis. In an ideal scenario, all events in a single measurement would thus be spatially well-localized around $x$. However, there are still a non-negligible portion of events with an interaction point away from the vicinity of $x$, e.g. gamma-rays that scatter at wide angles within the lead collimator. These events broaden the distribution of $\ln{S_1/S_0}$ and make the position reconstruction more inaccurate. This effect can be observed in Fig.~\ref{Fig:positionCut} for a \Na{} measurement at roughly the middle point ($x=\SI{24}{\cm}$). Here, the signal ratio is shown to be correctly centered at $\ln{S_1/S_0} \sim 0$, specially visible for events at the Compton-edge $\sqrt{S_0 \cdot S_1} \approx \SI{18}{\kilo\adc~\ns}$\footnote{The integrated charge in a waveform includes a multiplicative factor of \SI{4}{\ns} to account for the sample size of the digitizer. This can prove useful when comparing results from digitizers with different sampling sizes. This factor is removed later during the analysis of results.}. However these events are also widely distributed between $\ln{S_1/S_0} = \pm 1$. Such position effect provokes a broadening of the Compton-edge, making the fit of the distribution more challenging. We thus need to impose a position cut on $\ln{S_1/S_0}$ to ensure selected events with an interaction point at $x$ in the longitudinal axis of the bar. The selected region is bounded by the red lines in Fig.~\ref{Fig:positionCut}, which are defined as one standard deviation away from the mean signal ratio value. An example of the resulting Compton-scattered electron spectrum for \Na{}, after applying the position cut, and the corresponding fit can be observed in Fig.~\ref{Fig:comptonEdgeData}. 

\begin{figure}
    \centering
    \subfigure[Light-collection evolution]{\includegraphics[width=0.78\textwidth]{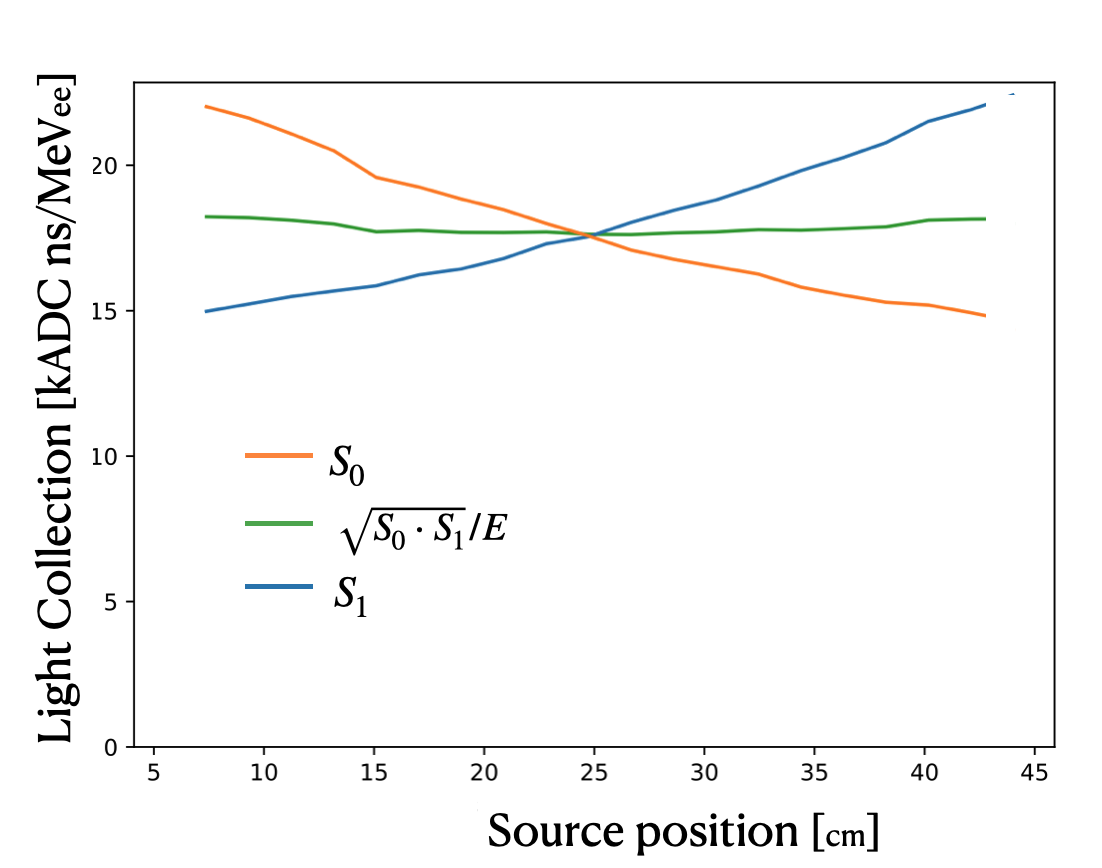} \label{Fig:AttCurves}} 
    \subfigure[Signal ratio evolution]{\includegraphics[width=0.78\textwidth]{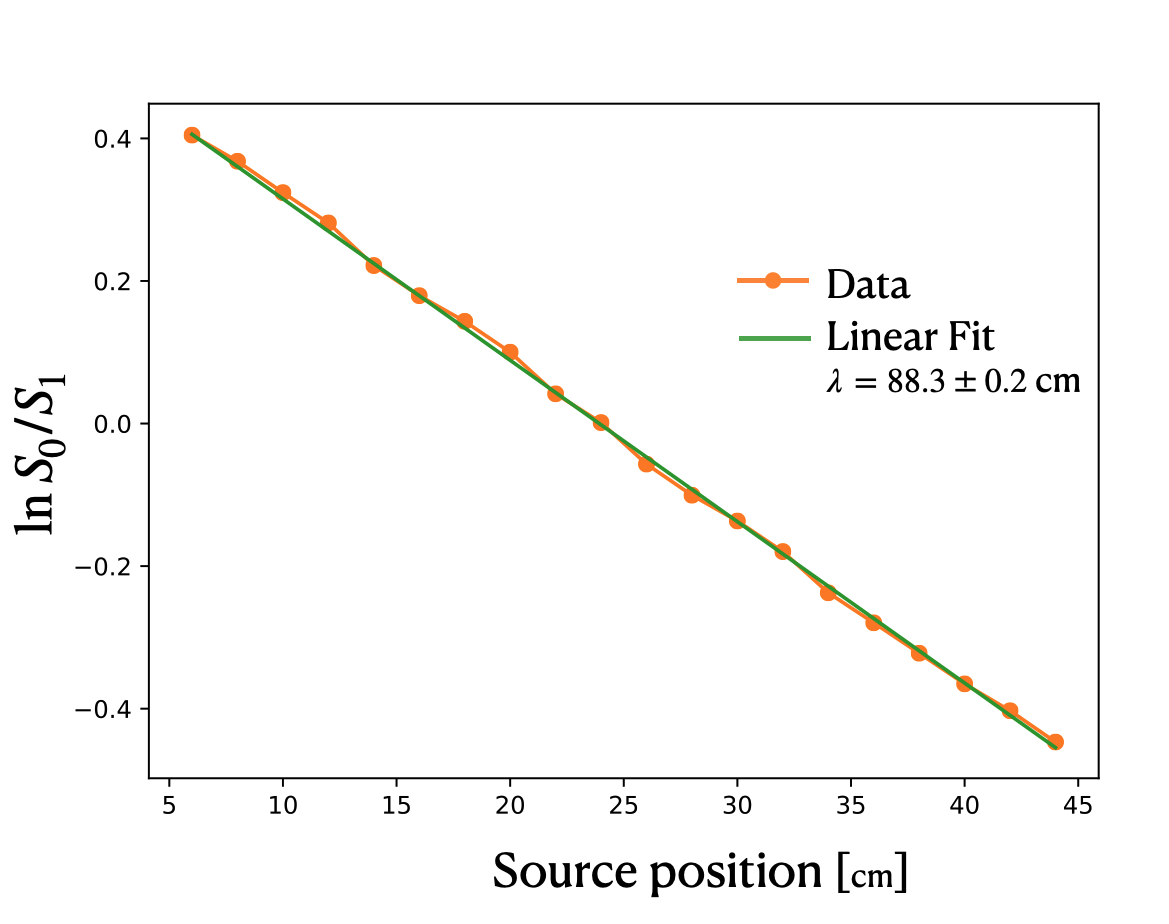} \label{Fig:AttCurve}} 
    \caption{Measured (a) light-collection and (b) signal-ratio for bar \#10 using \Na{} source across the longitudinal axis of bar.}
\end{figure}

This process can be performed over each individual \PMT{} signal, as well as for the geometric mean of the two. The relevant parameter obtained from the Compton-edge fit for this part of the study is $\varepsilon$. From it, we calculate the position of the Compton-edge and display its evolution in Fig.~\ref{Fig:AttCurves}. There it can be observed that individual \PMT{} signals evolve exponentially with $x$ while $\sqrt{S_0\cdot S_1}$ is position-independent, as described in Eqs.~\ref{Eq:S0}, \ref{Eq:S1} and \ref{Eq:sqrtS0S1}. The three lines displayed cross each other at the middle point ($x=L/2$), which demonstrates that both \PMT{}s are properly gain-matched, i.e $S_0 = S_1 = \sqrt{S_0 \cdot S_1}$. Fig.~\ref{Fig:AttCurve} shows the evolution of the signal ratio described in Eq.~\ref{Eq:lnS0S1}, together with a linear fit of the distribution from which can extract $\lambda$ for each bar. Finally, using Eq.~\ref{Eq:scale} with extracted $\lambda$ and $\varepsilon$, and using a \PMT{} gain of $G = \SI{30}{\adc\per\pe}$, we calculate $\ell$. Since this number is position-independent, we calculate the average for all 20 measurements.

\begin{figure}
    \centering
    \subfigure[Eff. attenuation length $\lambda$ distribution]{\includegraphics[width=0.78\textwidth]{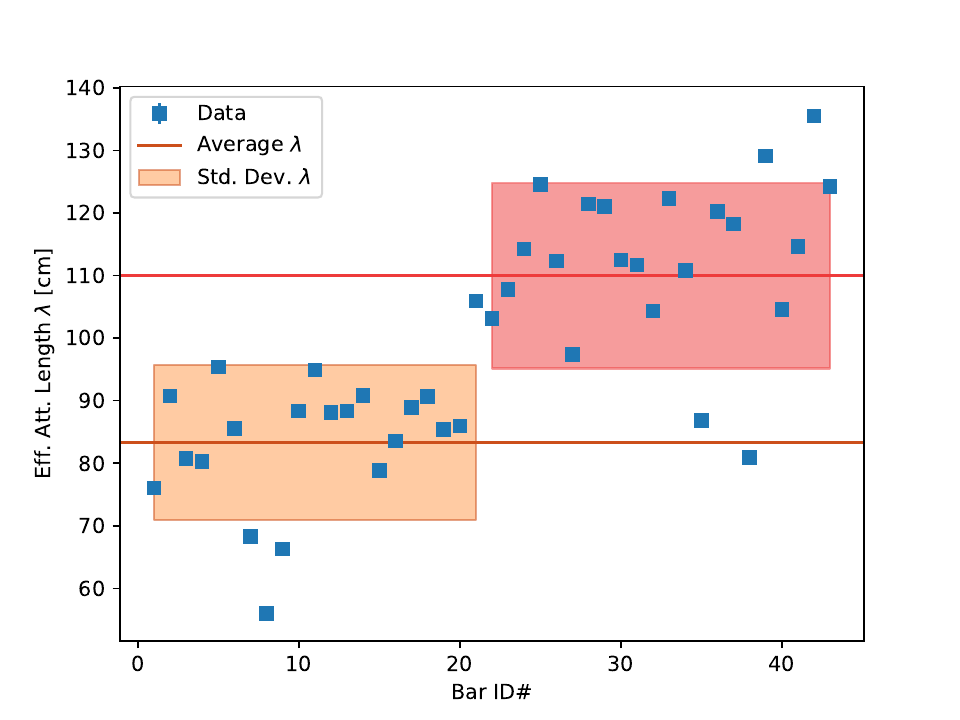}
    \label{Fig:AttLengthDist}}
    \subfigure[Eff. light output $\ell$ distribution]{\includegraphics[width=0.78\textwidth]{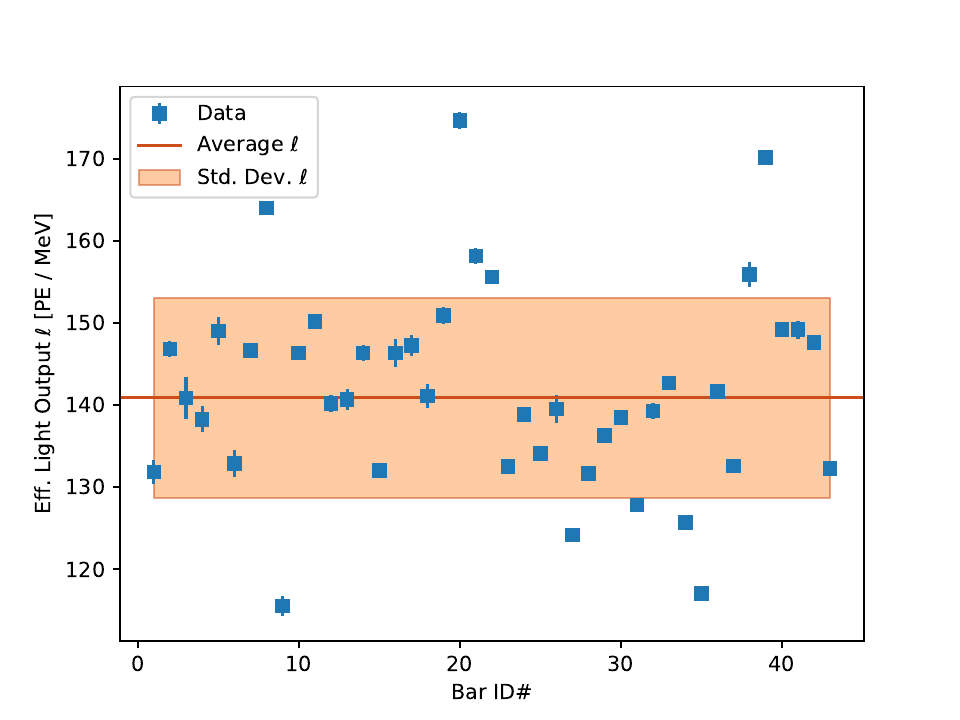}
    \label{Fig:LightOutputDist}}
    \caption{Measured \textit{effective} (a) attenuation length $\lambda$ and (b) light output $\ell$ distributions for the 44 \EJ{} scintillator bars tested. Two subgroups of bars can be distinguished with different average attenuation $\lambda_1$ and $\lambda_2$, marked by orange and red shades. These subgroups cannot be identified when looking at $\ell$ distribution.}
\end{figure}

Both analyses are thus performed for all \EJ{} bars:
\begin{enumerate}
    \item The resulting $\lambda$ distribution can be categorized into two subgroups with average values $\lambda_1 = \SI{83(12)}{\cm}$ and $\lambda_2 = \SI{110(14)}{\cm}$, shown in Fig.~\ref{Fig:AttLengthDist}. Throughout the bar production run, improvements to formulation and process were made, accounting for some of the variation observed. Generally, improved optical clarity was observed in later castings. 
    \item The distribution of $\ell$ for the 44 bars is shown in Fig.~\ref{Fig:LightOutputDist}. No substantial difference can be observed between production batches. The relative light output for \EJ{} in the \textit{bare} configuration is $\ell_{\text{bare}} = \SI{141(12)}{\pe\per\MeV}$. Similar measurement has been performed with some of the bars with complete wrapping and optical coupling. This setup, as explained in Sec.~\ref{Sec:ExperimentalDescription}, gives us an indication of the typical performance in a detector scenario, where effort is taken to transport as much  scintillation light to the \PMT{}s as possible. The results show an increase of $\sim 200\%$ in terms of light output, bringing the average to $\ell_{\text{wrapped}} = \SI{421(32)}{\pe\per\MeV}$. This value is approximately half of the average effective light output observed for standard EJ-200 \cite{Misc:EJ200} plastic bars of similar geometry in an identical \textit{wrapped} setup.
\end{enumerate}

\subsection{Energy Response in Long-Bars} \label{Sec:EnergyScale}
The selection of gamma-ray sources shown in Tab. \ref{Tab:Sources} is used to study the linearity of the energy response of the \EJ{} bars. In this study, each one of the sources is moved to the middle position ($L=\SI{25}{\cm}$) along the longitudinal axis of the bars. Bars have been set to the \textit{wrapped} configuration. The same process as in Sec.~\ref{Sec:AttLength} is used to ensure localized energy depositions and to fit and extract the corresponding Compton-edge for each source. Similarly, energy scale is calculated using Eq.~\ref{Eq:scale}. If the material has a linear response, $\varepsilon$ should be the same for all energy depositions listed in Tab.~\ref{Tab:Sources}, meaning a linear relationship between $E_{\text{dep}}$ and the detector response $\sqrt{S_0\cdot S_1}$. 

\begin{figure}
    \centering
    \includegraphics[width=0.85\textwidth]{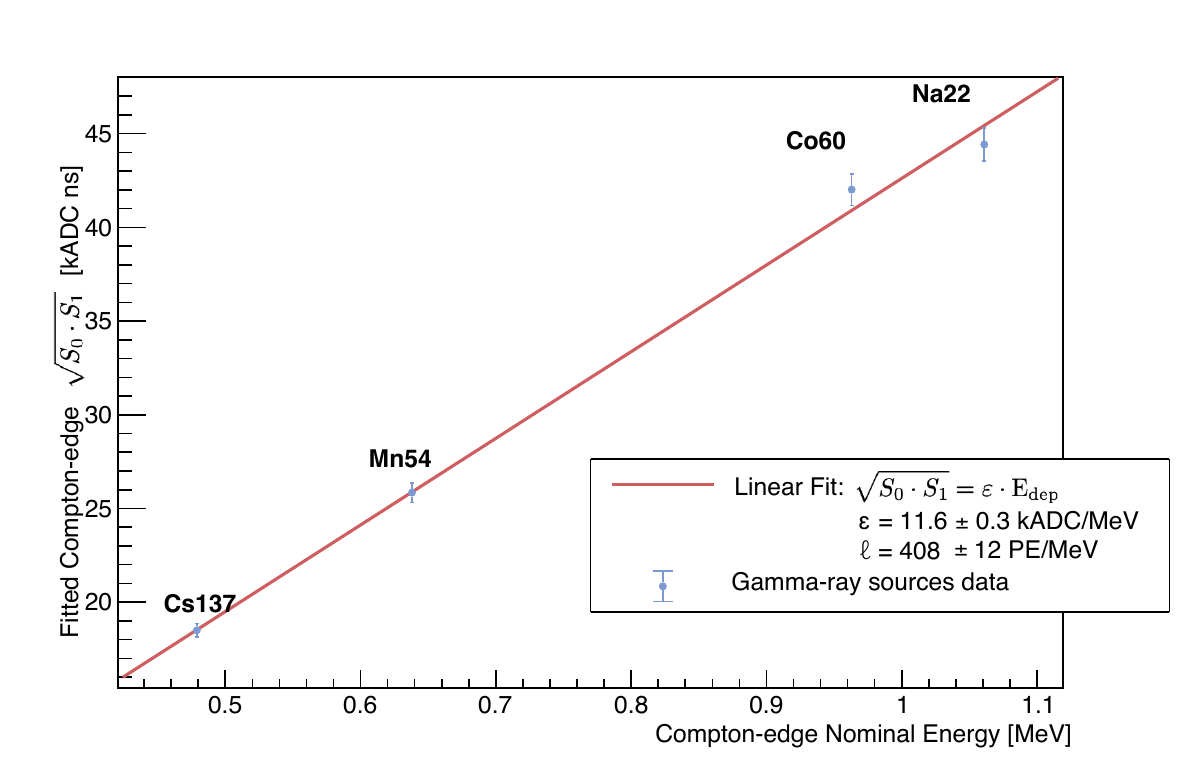} 
    \caption{Fitted Compton-edge position for gamma-ray sources listed in Tab.~\ref{Tab:Sources} using bar \#10. Sources have been placed at $x = L/2$ (middle point). Linear fit for all Compton-edge values, portraying a linear energy response.}
    \label{Fig:GammaLinear}
\end{figure}

The observed Compton-edges are plotted versus their nominal energies in Fig.~\ref{Fig:GammaLinear} for all sources. From the linear fit we obtain $\varepsilon = \SI{11.6 (0.3)}{\kilo\adc\per\MeV}$, which provides an alternative calculation of the \textit{effective} light output $\ell_{\text{wrapped}} = \SI{408 (25)}{\pe\per\MeV}$. This result is compatible within uncertainties with the valued obtaind in Sec.~\ref{Sec:AttLength}. This measurement demonstrates that the long-bar \EJ{} system has linear response at \unit{\MeV} scale energies. 

\begin{table} 
\begin{center}
\begin{tabular}{c | c | c}
Source & $E$[\unit{\MeV}] & $E_{\text{dep}}$ [\unit{\MeV}] \\
\hline \hline
\Cs{} & 0.667 & 0.479 \\
\Mn{} & 0.834 & 0.638 \\
\Co{} & 1.15 & 0.963 \\
\Co{} & 1.33 & 1.118 \\
\Na{} & 1.22 & 1.061
\end{tabular}
\caption{List of gamma-sources used during the test of energy respoonse's linearity for the \EJ{} bars. Second column depicts the nominal energy of the main gamma emission, while third column shows the energy value of the single-scattering Compton-edge.}
\label{Tab:Sources}
\end{center}
\end{table}

\subsection{Neutron Characterization for Long-Bars}
\label{Sec:NeutronStudies}
The neutron-identification performance of the \EJ{} plastics can be quantified through the \PSD{} parameter defined in Eq.~\ref{Eq:PSD} and the figure of merit \FoM{} between \PSD{} regions, defined in Eq.~\ref{Eq:FoM}. For this purpose, the response of the bars to a \Cf{} gamma-neutron source is used, as described in Sec.~\ref{Sec:ExperimentalDescription}. The bars are prepared in the \textit{wrapped} configuration  to optimize \PSD{} band separation. Each measurement is 1-hour long with an additional 1-hour of background data taken after removing the source. This background is later subtracted from the foreground. After acquisition, integrated charge is transformed into reconstructed energy by means of an energy scale database, created previously for each bar from the multi-position gamma-ray measurements and the calculation of $\varepsilon$ described in Sec.~\ref{Sec:AttLength}.

\begin{figure}[t]
    \centering
        \includegraphics[width=0.85\textwidth]{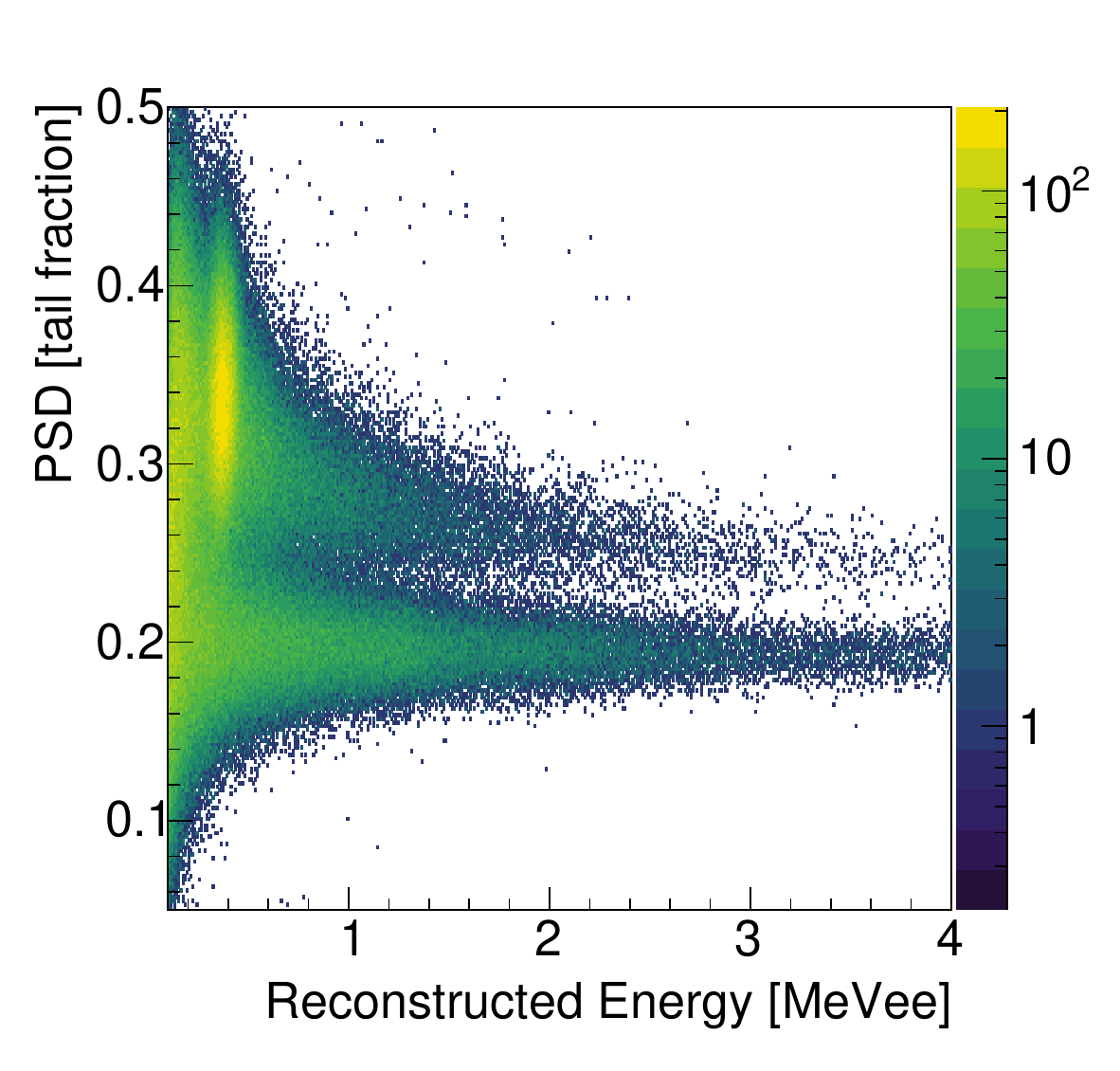}
        \caption{\PSD{} parameter in terms of reconstructed energy. Two main bands can be observed for nuclear and electron-like recoils respectively. neutron-captures in \Li{} appear prominently in the upper \PSD{} band at about \SI{0.4}{\MeVee} reconstructed energy and \PSD{} between \numlist{0.28;0.42}.}
        \label{Fig:MoneyPlot}
\end{figure}

The distribution of reconstructed energy versus \PSD{} is shown in Fig.~\ref{Fig:MoneyPlot} for the \Cf{} data. In this plot, three main regions can be identified: (1) an electronic-recoil band for \PSD{} values below $\sim0.24$, generated by gamma-rays interacting with electrons in the material; (2) a nuclear-recoil band for \PSD{} values above $\sim0.24$, generated by fast-neutrons scattering with nuclei; and (3) a small region within the nuclear-recoil band at energies \SI{\sim 0.4}{\MeVee}, generated by thermal-neutron-captures in \Li{} isotopes and colloquially referred as \textit{capture island}. The exact position and width of both bands have been calculated by approximating each band's projection ($i=\gamma,n$) into the \PSD{} axis to a Gaussian function, parameterized by a mean ($\mu_{i}$) and a standard deviation ($\sigma_{i}$). The capture island is fitted to a two-dimensional Gaussian function parameterized by $E_{\text{Li}}, \PSD{}_{\text{Li}}$ and $\sigma_{E,\text{Li}}, \sigma_{\PSD{},\text{Li}}$, means and standard deviations in the energy and \PSD{} axes respectively. 

In our studies two \FoM{} values are calculated: one between fast-neutron and gamma bands in the $\SIrange{1}{2}{\MeVee}$ region ($\FoM{}_{n}$), and another one between the capture island and gamma band in the [$E_{\text{Li}}- \sigma_{\text{Li}}$, $E_{\text{Li}} + \sigma_{\text{Li}}$] region ($\FoM{}_{\text{Li}}$). The results show similar discriminating power for both regions with $\FoM{}_{n} = 1.27\pm 0.11$ and $\FoM{}_{\text{Li}} = 1.25\pm 0.18$. These numbers are lower than the ones obtained in Sec.~\ref{Sec:Small}, which is expected given the larger volume of the bars, and thus the lower light collection efficiency.  \\

Efficient \IBD{} detection using \Li{}-doped material emphasizes captures on that agent as opposed to other material constituents. Absolute neutron-capture-efficiency is hard to measure reliably, since a highly detailed understanding of the detector setup, its environment, and the event selection are required. Such a measurement would be beyond the scope of this work, so a comparison of the effective capture-efficiency in \Li{} between data and simulation is performed. Good agreement would allow us to provide an estimate of the absolute capture-efficiency using truth information from simulations. 

The capture-efficiency in \Li{} depends on the remaining potential candidates for thermal-neutron capture in the material. If we exclude the specific loading material (i.e. \Li{}), the largest capture cross-section in organic scintillators is for \Hy{} nuclei. This type of captures results in the emission of a \SI{2.2}{\MeV} de-excitation gamma-ray by the capturing nuclei, a type of event that falls within the electronic-recoil band in Fig.~\ref{Fig:MoneyPlot}. These events strongly compete with background in the \PSD{}-energy space and thus are generally discarded for \IBD{} coincidence search in favor of a more specific event selection provided by captures in \Li{}. Thus, to determine the neutron-capture-efficiency we compare the amount of \Li{} captures versus captures in \Hy{}. Simultaneously, a Geant4 \cite{G4} simulation of the setup and data acquisition is performed and analysed together with the bench-top measurements, allowing us to estimate how well neutron dynamics are represented in our simulation model of these plastics. The single bar used for this study is configured in \textit{bare} conditions to facilitate a straightforward geometry for the simulation model.

Selection of thermal-neutron-captures requires a coincidence search throughout all the events. A coincident prompt-delayed pair is found when both prompt and delayed-events fulfill \PSD{}, energy and time conditions. The selection process is:
\begin{enumerate}
    \item Scan all recorded events for potential capture delayed-events on \Li{} or \Hy{} according to Tab.\ref{Tab:SelectionCuts}. This scan prioritizes tagging captures in \Li{} before \Hy{}.
    \item Once a delayed-event candidate is found, all events in the range of \SIrange{-250}{0}{\micro\second} happening prior to it are scanned for a prompt-event candidate according to Tab.\ref{Tab:SelectionCuts}.
    \item If a prompt-event is found, then all events in the range of \SIrange{-3000}{-250}{\micro\second} happening prior to the delayed-event are scanned. Any coincidences found in this range will be treated as accidental, and will be used for accidental subtraction later. Since this window is larger than the coincidence event window, the accidental count is down-scaled prior to the subtraction.
\end{enumerate}

\begin{table}
\caption{\PSD{} and energy selection cuts for \Li{} and \Hy{} thermal-neutron-capture prompt-delayed coincidences.}
\label{Tab:SelectionCuts}
\begin{center}
\begin{tabular}{c | c | c}
  & \textbf{\PSD{} cut} & \textbf{Energy cut} \\
\hline \hline
\textbf{Prompt - H} & [n-band - $1\sigma$, 0.4] & [E$_{\text{Li}} + 1\sigma$, 10]\unit{\MeV} \\
\textbf{Delayed - H} & [0, gamma-band + $3\sigma$] & [0.1, 1.6]\unit{\MeV}\\
\hline
\textbf{Prompt - \Li{}} &  [n-band - $1\sigma$, 0.4] & [E$_{\text{Li}} + 1\sigma$, 10]\unit{\MeV} \\
\textbf{Delayed - \Li{}} & \PSD{}$_{\text{Li}} \pm 1\sigma$ & E$_{\text{Li}} \pm 1\sigma$ \\
\end{tabular}
\end{center}
\end{table}

The delayed-event selection cuts in Tab.~\ref{Tab:SelectionCuts} are configured for two different type of energy depositions following the thermal-neutron-capture in \Hy{} and \Li{}:
\begin{enumerate}
    \item Captures in \Hy{} are defined by the emission of a de-excitation \SI{2.2}{\MeV} gamma-ray by the resulting deuterium nuclei after the neutron-capture. The energy deposition for these events is limited to \SI{1.6}{\MeV} since that is the maximum energy that can be deposited in the material before the gamma-ray Compton-scatters out of the volume. 
    \item Captures in \Li{} are the result of the ionization produced by an $\alpha$-particle plus tritium nuclei, products from the neutron-capture on \Li{}. These ions carry a combined kinetic energy of \SI{4.78}{\MeV} that results in a compact mono-energetic deposition through a series of ion-nucleus inelastic scatterings. The reconstructed energy can be found around the \SI{0.4}{\MeVee} mark due to quenching, while the \PSD{} parameter for these events is in the nuclear-recoil band. 
\end{enumerate}

\begin{figure}
    \centering
    \subfigure[prompt-events for coincidences in H]{\includegraphics[width=0.48\textwidth]{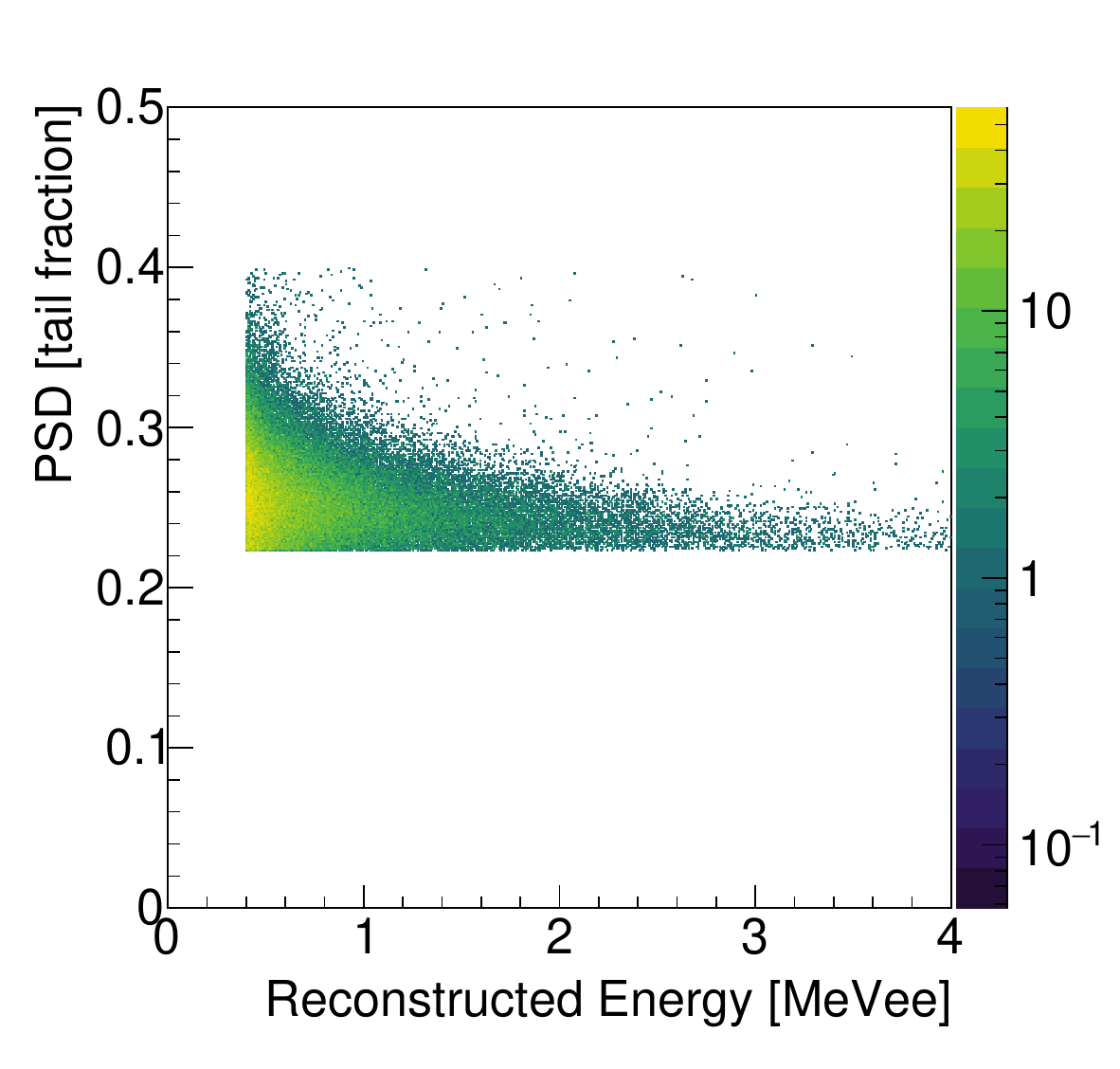}} 
    \subfigure[delayed-events for coincidences in H]{\includegraphics[width=0.48\textwidth]{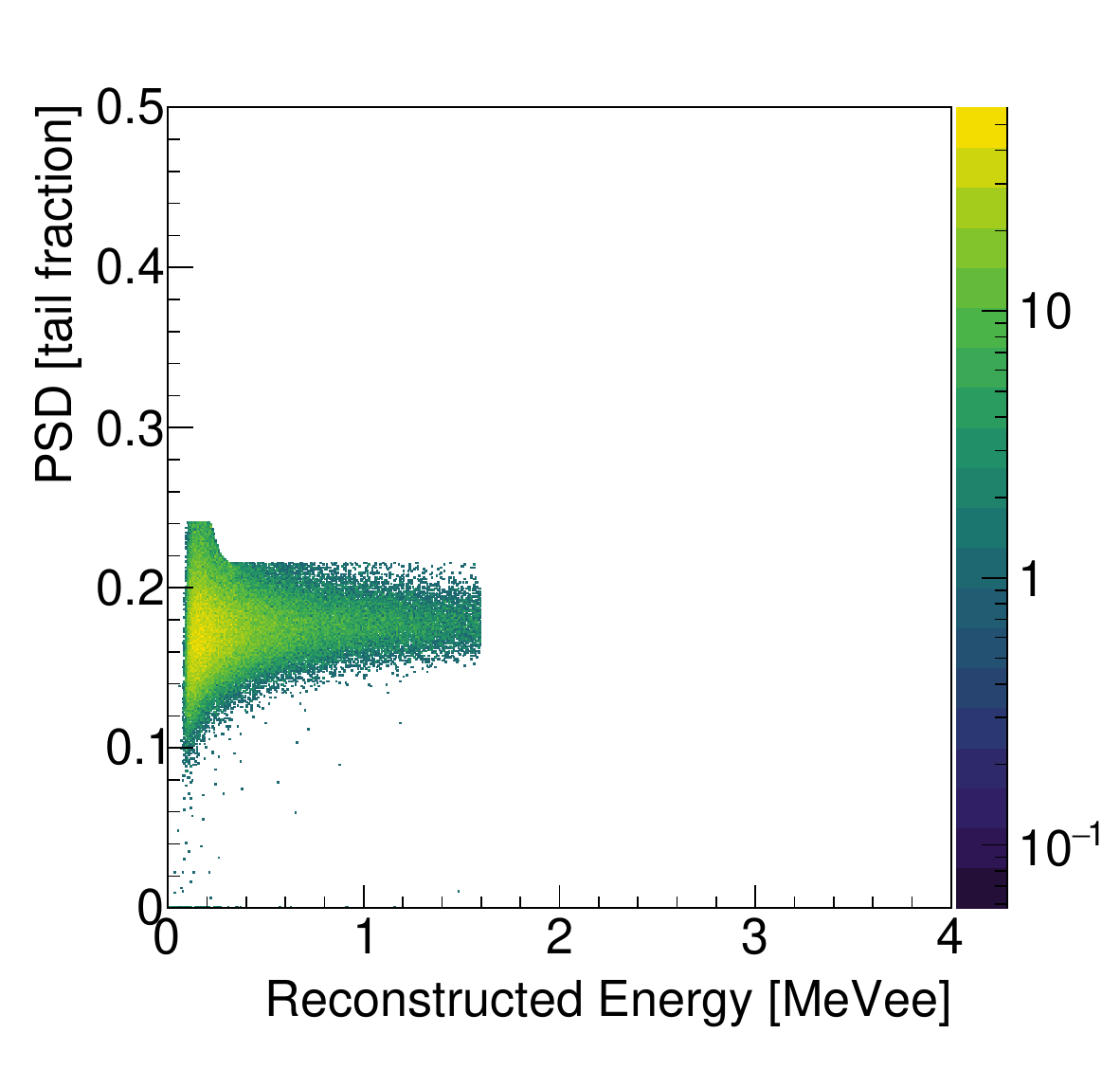}} 
    \subfigure[prompt-events for coincidences in \Li]{\includegraphics[width=0.48\textwidth]{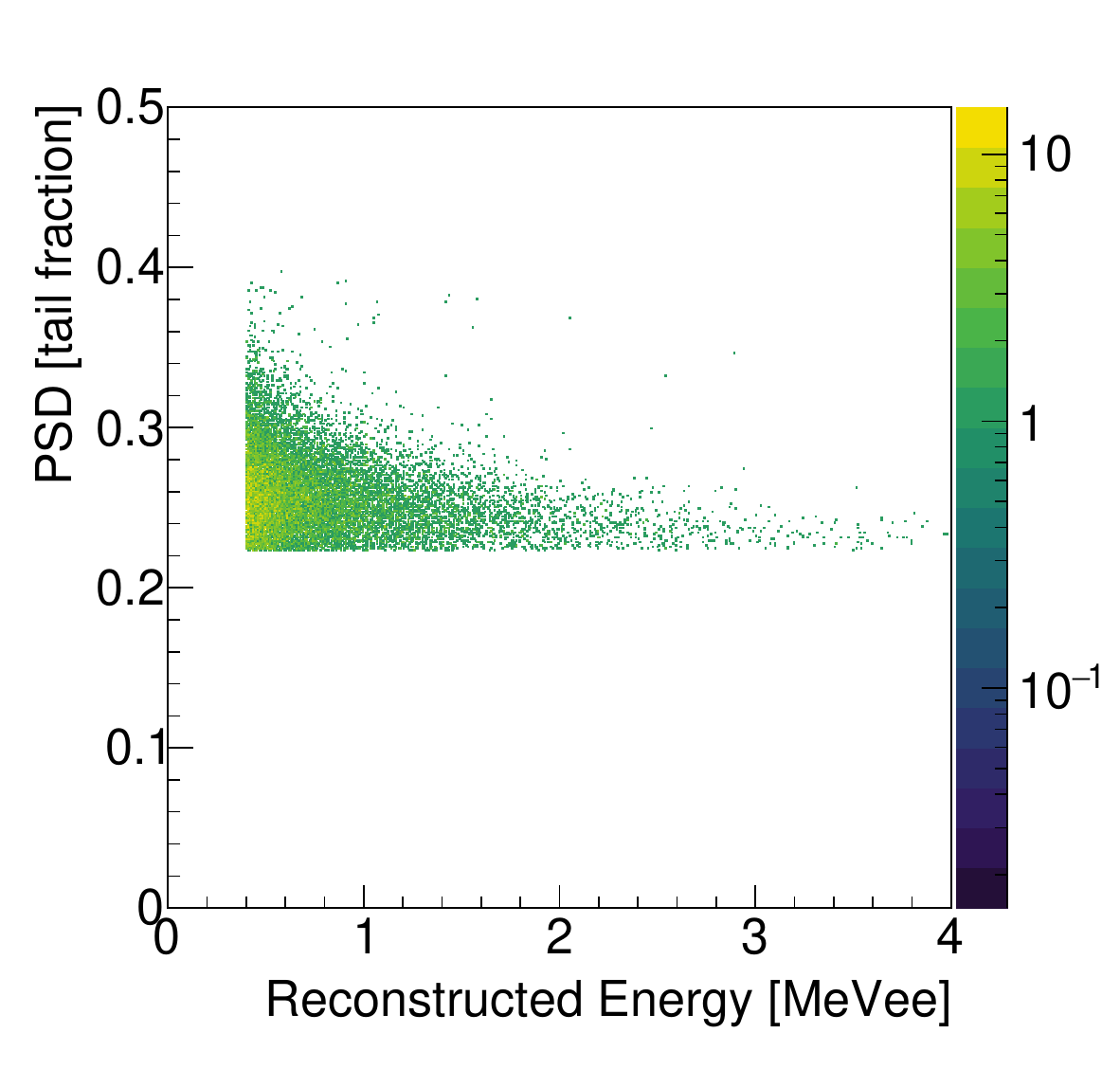}}
    \subfigure[delayed-events for coincidences in \Li]{\includegraphics[width=0.48\textwidth]{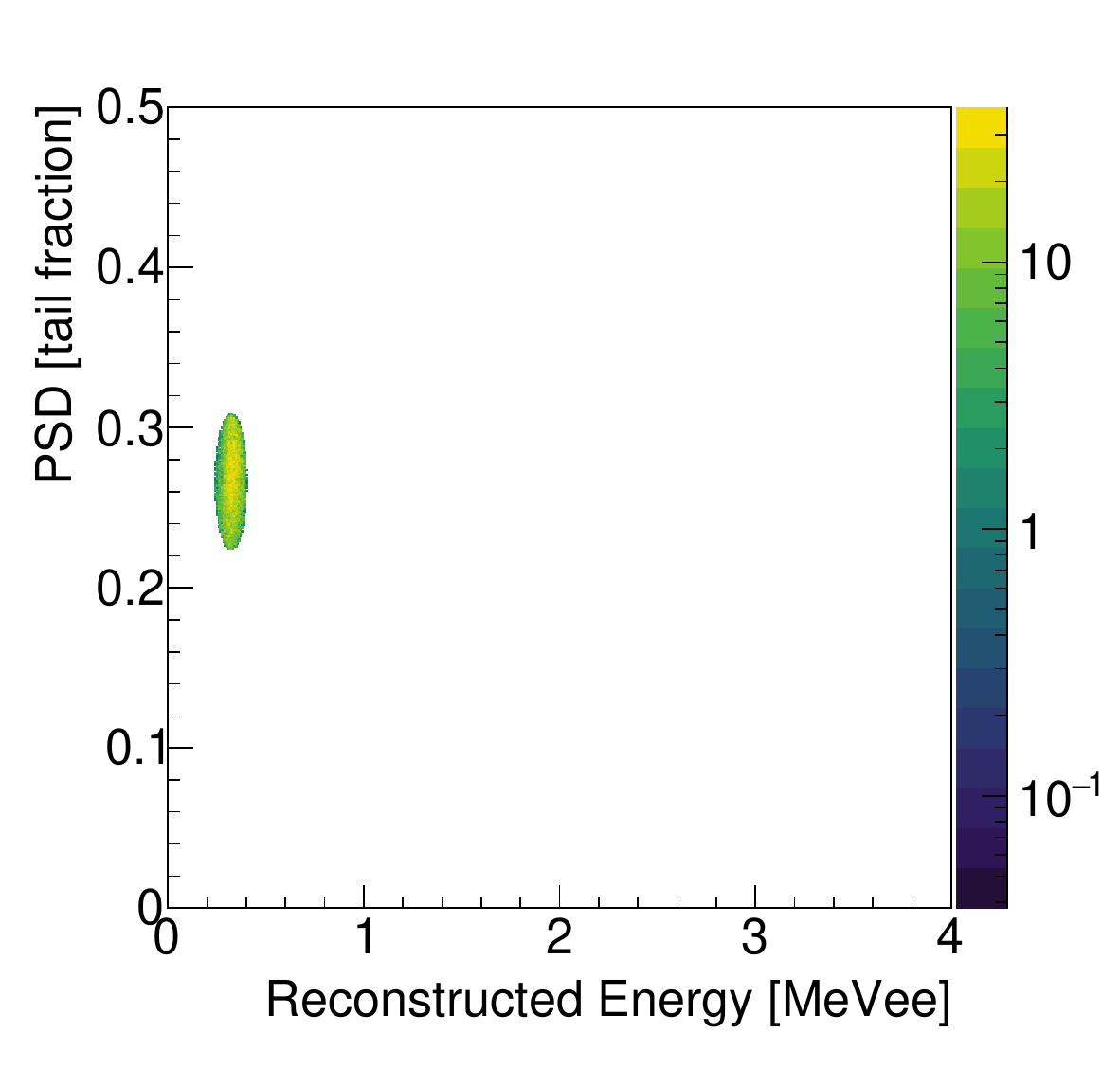}}
    \caption{\PSD{} vs reconstructed energy distribution of events after applying selection cuts in Tab.~\ref{Tab:SelectionCuts} on \Cf{} data. Fast-neutron-induced nuclear recoils are selected as prompt-events, while thermal-captures in \Li{} and \Hy{} are chosen as delayed-events.}
    \label{Fig:SelectionCuts}
\end{figure}

The selection of prompt-events is done identically for both types of coincidences, and it is primarily defined by nuclear-recoils produced by an incoming fast-neutron from the \Cf{} source before thermalizing and capturing. The energy spectrum for these fast-neutrons approximately ranges between $\SIrange{0.5}{10}{\MeVee}$ \cite{Cf252Spectrum}. To avoid mis-selecting captures in \Li{} as prompts, the low-energy end for the prompt selection range is set $1\sigma$ away from the capture island center while the high-energy end stretches until the end of the spectrum. \PSD{} for this type of events lies within the nuclear-recoil band, and thus we use $1\sigma$ below the centroid as low-\PSD{} end of the selection, while opening the high-\PSD{} end all the way until 0.4. \\ 

The position of the selection cuts in Tab.~\ref{Tab:SelectionCuts} substantially impacts the measured capture-fraction. For this reason, the cuts are defined relative to the position and width of both recoil bands and the thermal-neutron-capture island. This way, cuts affect the same topology of events in data and simulation. The events in data surviving each set of selection cuts are depicted in Fig.~\ref{Fig:SelectionCuts}.

\begin{figure}
    \centering
    \subfigure[Coincidences in MC \Li{}]{\includegraphics[width=0.48\textwidth]{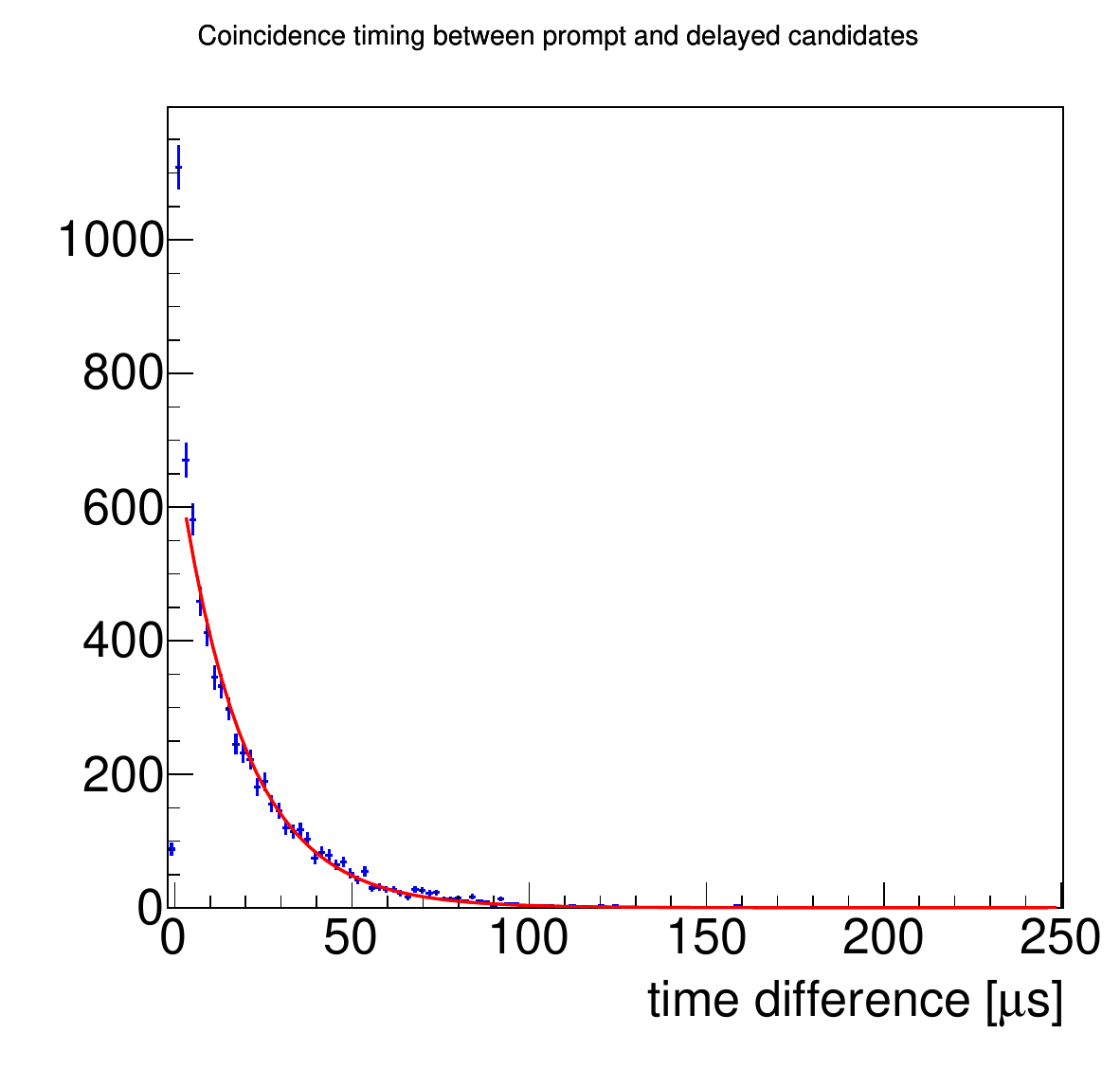}} 
    \subfigure[Coincidences in MC \Hy{}]{\includegraphics[width=0.48\textwidth]{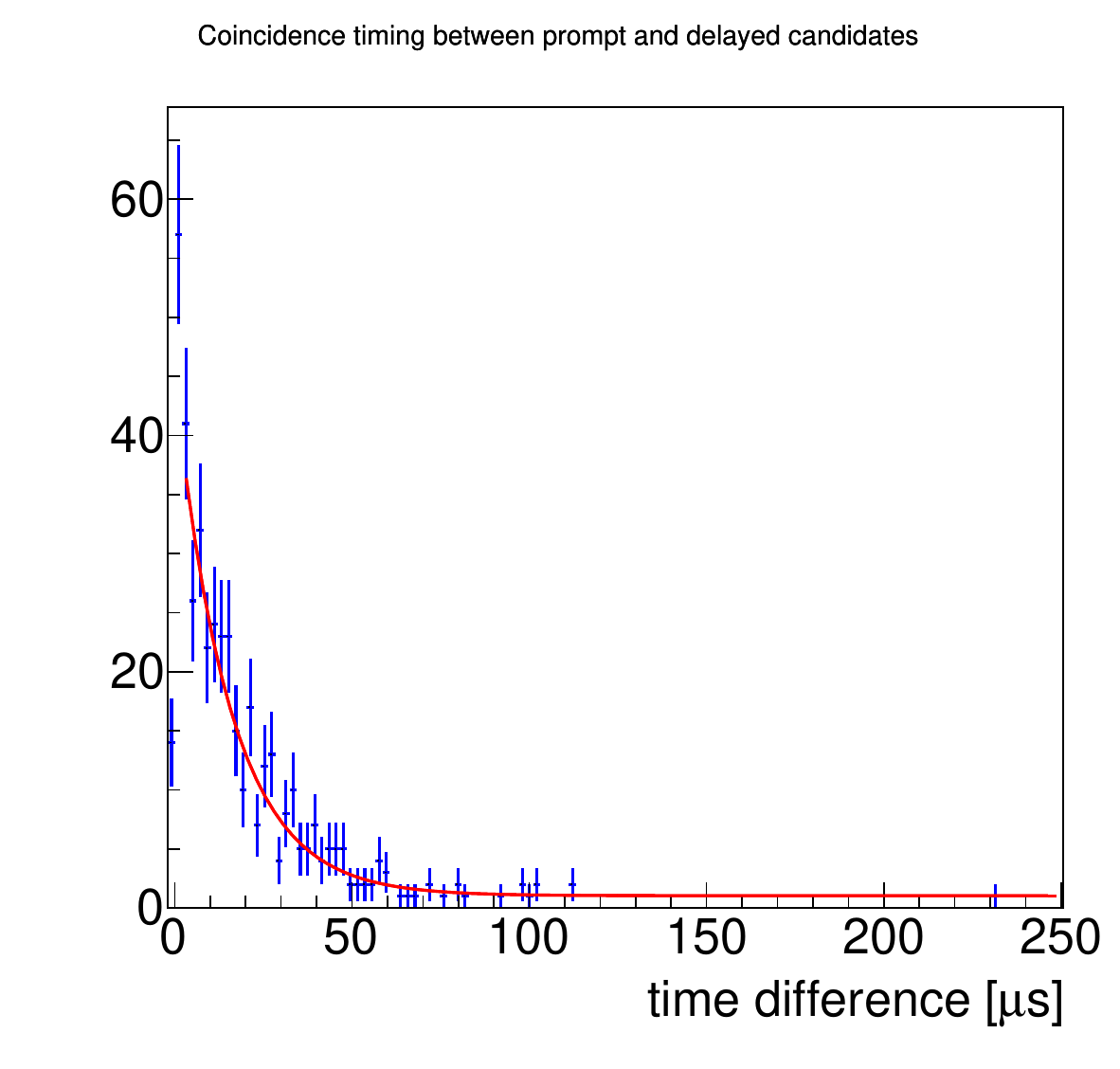}} 
    \\
    \subfigure[Coincidences in data \Li{}]{\includegraphics[width=0.48\textwidth]{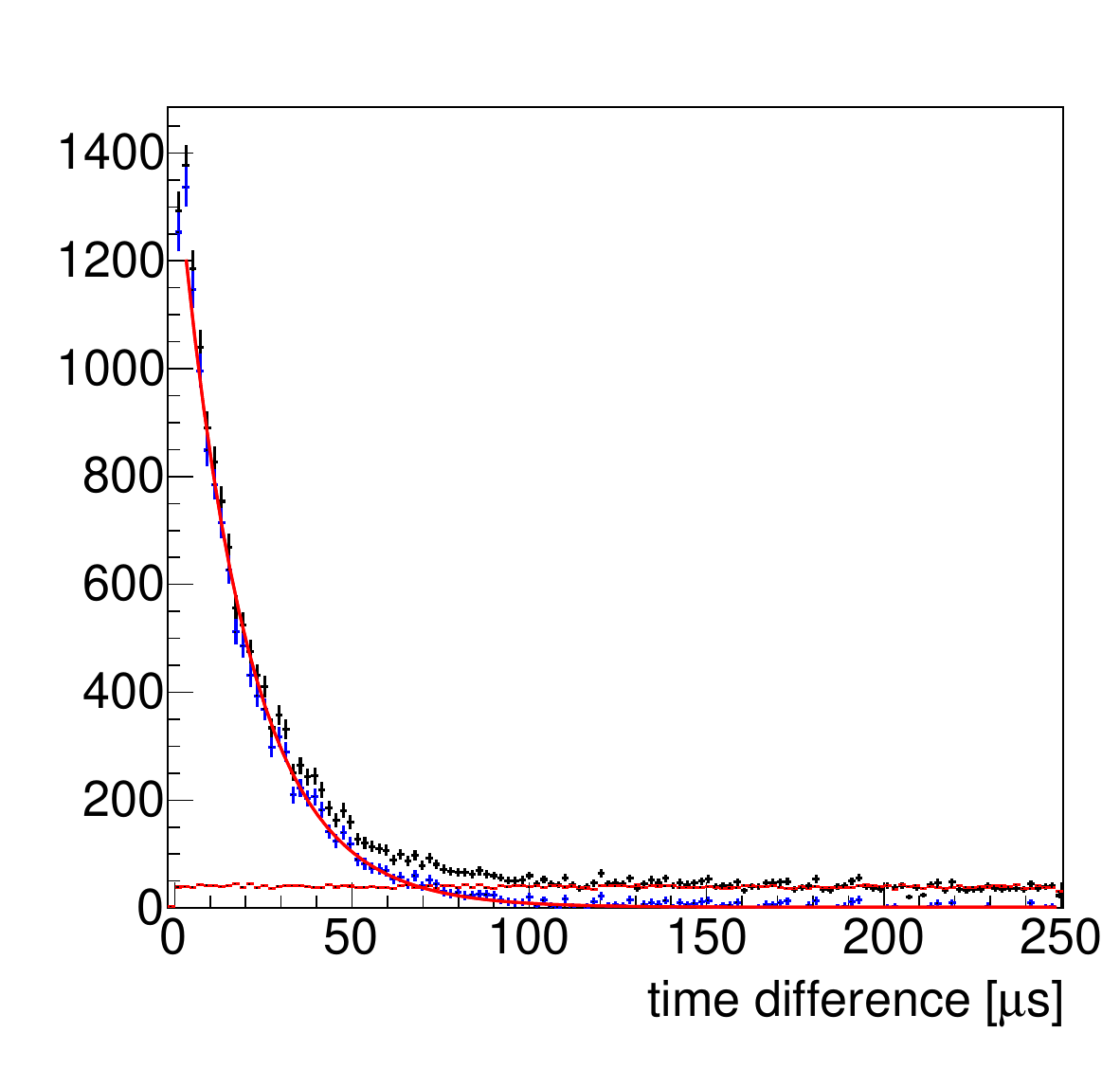}}
    \subfigure[Coincidences in data \Hy{}]{\includegraphics[width=0.48\textwidth]{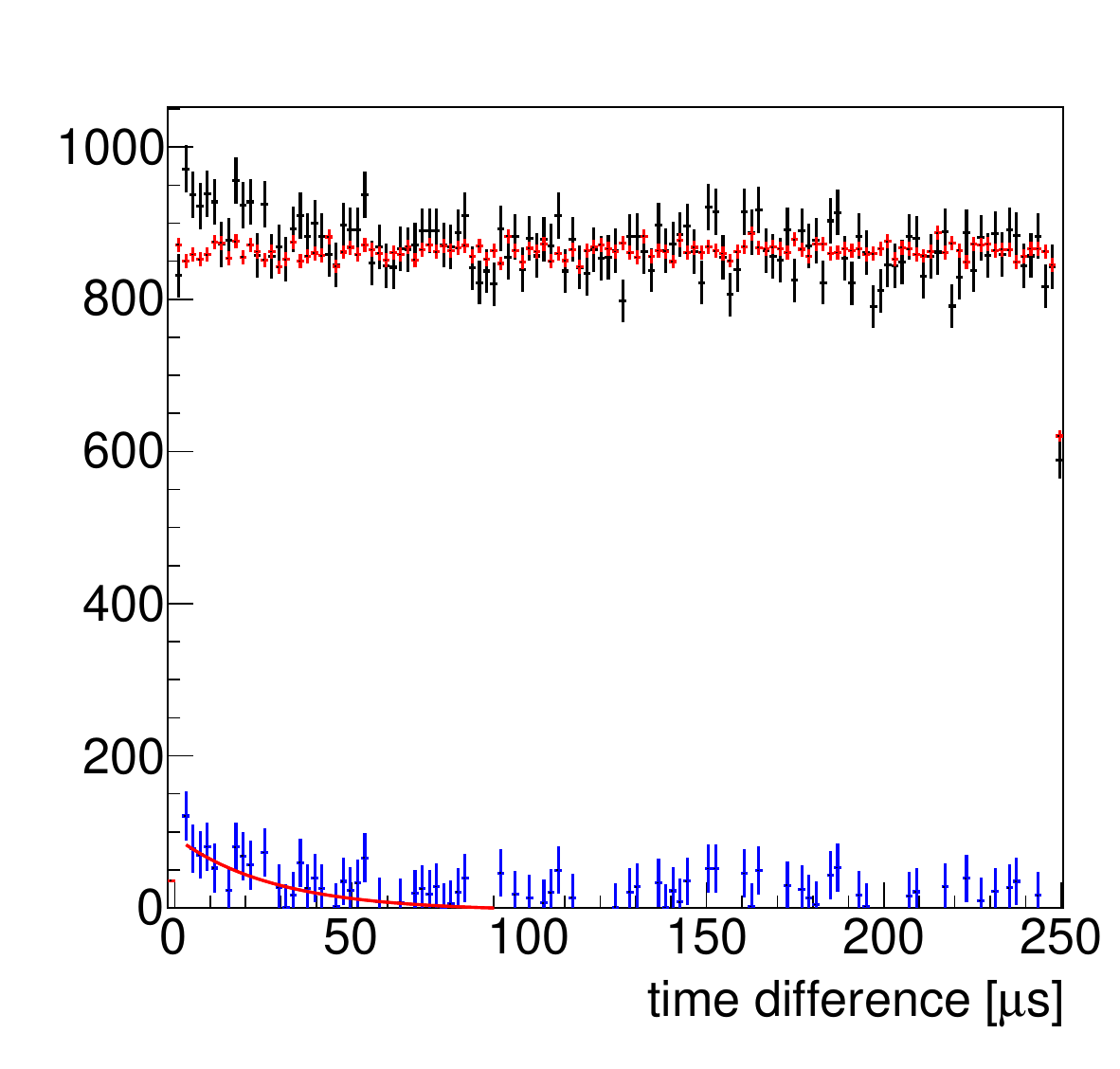}}
    \caption{Coincidence time distributions for data and MC datasets after applying corresponding selection times for captures in both \Hy{} and \Li{}.}
    \label{Fig:CoinTimes}
\end{figure}

The same events can be plotted in terms of the coincidence time, i.e the time difference between the prompt-delayed pair. Such distribution for captures in \Hy{} and \Li{} is shown in Fig.~\ref{Fig:CoinTimes}. The bottom figures showcase the time difference distribution measured for data (black dots) together with the estimated accidental contribution (red dots). The subtracted coincidence distribution is depicted in blue, and can be approximated to an exponential function:
\begin{equation}
    N_{corr} = A\exp{(-t/\tau)} + c,
\end{equation}
with $c\sim0$ when accidental coincidences are properly subtracted, and $\tau$ being the mean capture-time in a specific isotope and material. Our results show an excellent agreement between data and simulation for \Li{} with $\tau_{\text{Li,dat}} = $\SI{15.1(0.6)}{\micro\second} and $\tau_{\text{Li,sim}} = $~\SI{14.7(0.8)}{\micro\second}. Measurement of the capture-time for \Hy{} proved to be more challenging as rate of correlated events is much smaller than the accidental rate. Data and simulation capture-times for \Hy{} are $\tau_{\text{H,dat}} = \SI{8.4(6.7)}{\micro\second}$ and $\tau_{\text{H,sim}} =\SI{14.6(3.0)}{\micro\second}$ respectively. These results are compatible with each other but present large uncertainties and show that the current composition enhances captures in \Li{} which is the desired outcome. The capture-fraction numbers are also found to be compatible for data and simulation: $f_{\text{dat}} = 16.8\pm2.4$ and $f_{\text{sim}} = 16.5\pm2.3$.

As mentioned previously, capture-fraction and capture-time calculations are heavily dependent on the selection cuts and can vary largely depending the energy and \PSD{} limits. These results should not be understood thus as an absolute measure of neutron-capture statistics, given that selection and geometry effects have not been quantified. However, an absolute capture-efficiency can be extrapolated to data from true information in the simulation, provided the proven compatibility between both.

To avoid geometry effects from using a single plastic bar, a simulation of a large and symmetric volume of \EJ{} has been performed. A plastic scintillator sphere of \SI{1}{\m} radius is simulated together with a \SI{10}{\eV} neutron particle-gun at its center. This configuration has been chosen to ensure complete capture of all thermal-neutrons within the active volume. True information about neutron-captures is recorded and categorized according to the capturing isotope. The results show an absolute efficiency of captures in \Li{} of $85\%$ with a mean thermalization path of \SI{41.75}{\milli\meter} and capture-time of $\tau_{\text{Li,true}} = ~\SI{37.2}{\micro\second}$. The rest of captures happen in \Hy{} with a $13\%$ efficiency and $\tau_{\text{H,true}} = \SI{36.2}{\micro\second}$, or in other isotopes like C and N at a $\sim 2\%$ rate.

\subsection{Material Stability} \label{Sec:Stability}

Previous studies \cite{Zaitseva:2018} have found that \PPO{}-based \PSD{}-capable plastic scintillators can suffer from gradual degradation of their properties over time due to the \PPO{} dye included at high concentration to provide \PSD{} capabilities. A recurring problem in these type of scintillators is the outgassing (leaching) of \PPO{}, which can crystallize on the machined scintillator surfaces degrading light collection.

This effect has been also observed on the \EJ{} bars measured in this study. When the bars are stored in the \textit{bare} configuration, little to no leaching is observed, probably because outgassed \PPO{} is mixed with a much larger air volume or/and leachability being reduced on material surface by oxidation, as evidenced by yellowing. However, when the bars are left in \textit{wrapped} condition for weeks or more, \PPO{} leachate can appear between the wrapping and the bars, as the outgassing is retained in the air pocket established for \textsc{tir} light transport. Regardless, the original performance is recovered if the precipitate is removed, which can be accomplished by a wipe cleaning with ethanol. The amount of \PPO{} leaving the material is small, and does not appear to affect the inherent material performance. Methods for reducing or eliminating this phenomenon are under investigation, including film surface barrier layers and/or reducing the material \PPO{} fraction. 

% \begin{figure}[h] 
%     \centering
%         \includegraphics[width=0.7\linewidth]{img/leaching.png}
%         \label{Fig:leaching}
%         \caption{Bar \#15 after taking measurements for some days in \textit{wrapped} configuration. The crystallization of \PPO{} can be observed on the surface of the plastic and on the reflective wrapping.}
% \end{figure}

\begin{figure}[h] 
    \centering
        \includegraphics[width=0.9\linewidth]{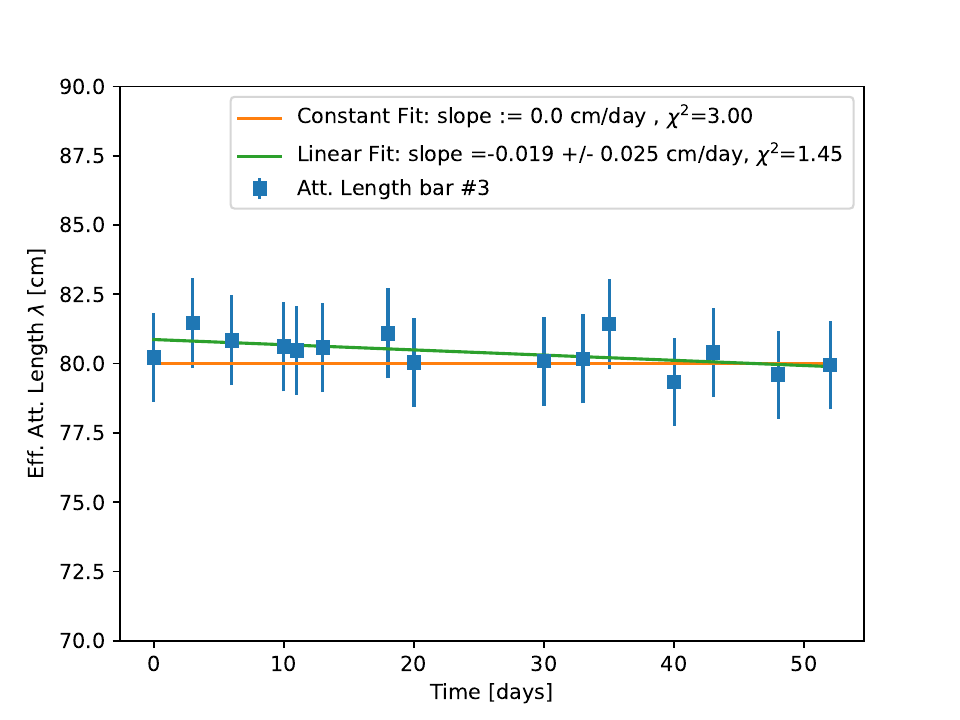}
        \caption{Evolution of attenuation length for bar \#3 in \textit{bare} conditions throughout the measurement campaign. Those 15 measurements correspond to approximately two months of data taking.}
        \label{Fig:att3}
\end{figure}

During the characterization of the bars described in Sec.~\ref{Sec:AttLength}, one of the four bars tested at any single time is set to be the same for every measurement. In particular, bar \#3 is chosen as the benchmark bar. After each set of measurements, the bar is extracted from the setup, cleaned up and reintroduced together with the new batch of plastics. Eff. attenuation length for this specific bar is thus tracked over approximately two months of measurements. This monitoring of bar \#3 is displayed in Fig.~\ref{Fig:att3}, showing a relatively stable trend over data-taking time. However, results in Fig.~\ref{Fig:att3} cannot be used to draw definitive conclusions about the long-term stability of the plastics, as the time frame is relatively short with respect to the operational expectations.   

Regardless, a $\chi^2$ analysis has been performed over bar \#3 data to test a degradation hypothesis over the null case where data is stable. Both a linear and constant fit have been performed for the data and shown in Fig.~\ref{Fig:att3}. $\chi^2$ values are calculated for both models. The relative $\Delta\chi^2 = 1.55$ corresponds to only one degree of freedom difference between hypotheses, which yields a $p$-value of $0.021$. This result showcases that both constant and linear models are compatible and that any discrepancy with a constant distribution is not statistically significant. Additionally, the slope obtained from the linear fit is compatible with 0 within the uncertainty of the fit, which confirms our conclusions. \\ 
% A more comprehensive study of the performance evolution for \EJ{} using a different setup and methodology is described in Sec.~\ref{Sec:Aging}.
% \subsubsection{Accelerated Aging Tests} \label{Sec:Aging}
The performance of \EJ{} over time was evaluated through a dedicated long-term testing of two variants of the plastic: ``Original'' formulation, representing the main formulation studied throughout this manuscript; and ``Current'' formulation, modified to increase the performance stability of the scintillators. This study was performed using the \textit{small} configuration, where the 2-inch plastic samples were left in different background conditions for varying amounts of time. 
% A complete list of the conditions applied to the plastics can be read in Tab.~\ref{Tab:testSeq}.
% \begin{table*}[t!]
% \caption{Testing sequence for accelerated aging measurements of \EJ{} samples in 2 groups: stored in air and in N$_2$. RT = room temperature storage (\SI{\simto 20}{\degreeCelsius}.} 
% \label{Tab:testSeq}
% \centering
% \sffamily
% \begin{tabular}{l|l|l}
%   \multicolumn{3}{ c }{\textbf{GROUP 1 (Air)}} \\
%   \hline
%   \textbf{Condition} & \textbf{Duration} & \textbf{Total Duration }  \\ 
%   \hline \hline
% Initial Measurement & - & - \\ \hline
% Air at \SI{60}{\degreeCelsius}  & 10 days & 10 days \\ \hline
% Air at RT  & 3 weeks & 4.5 weeks  \\ \hline
% Air at RT  & 4 weeks & 8.5 weeks  \\ \hline
% Air at RT  & 8.5 weeks & 17 weeks \\ \hline
% Air at RT  & 2 weeks & 19 weeks   \\ \hline
%  \multicolumn{3}{ c }{ } \\
%  \multicolumn{3}{ c }{\textbf{GROUP 2 (N$_2$ )}} \\
%  \hline
%  \textbf{Condition} & \textbf{Duration} & \textbf{Total Duration}  \\
%  \hline \hline
% Initial Measurement   & - & -     \\ \hline
% N$_2$ at RT   & 3 weeks & 3 weeks     \\ \hline
% N$_2$ at RT   & 6 weeks & 9 weeks     \\ \hline
% N$_2$ at \SI{60}{\degreeCelsius}   & 8 days & 10 weeks     \\ \hline
% N$_2$ at RT   & 7 weeks & 17 weeks     \\ \hline
% N$_2$ at RT   & 2 weeks & 19 weeks     \\ \hline
 
%  \end{tabular}
% \end{table*}

\begin{figure}[H]
    \centering
    \includegraphics[width=0.9\textwidth]{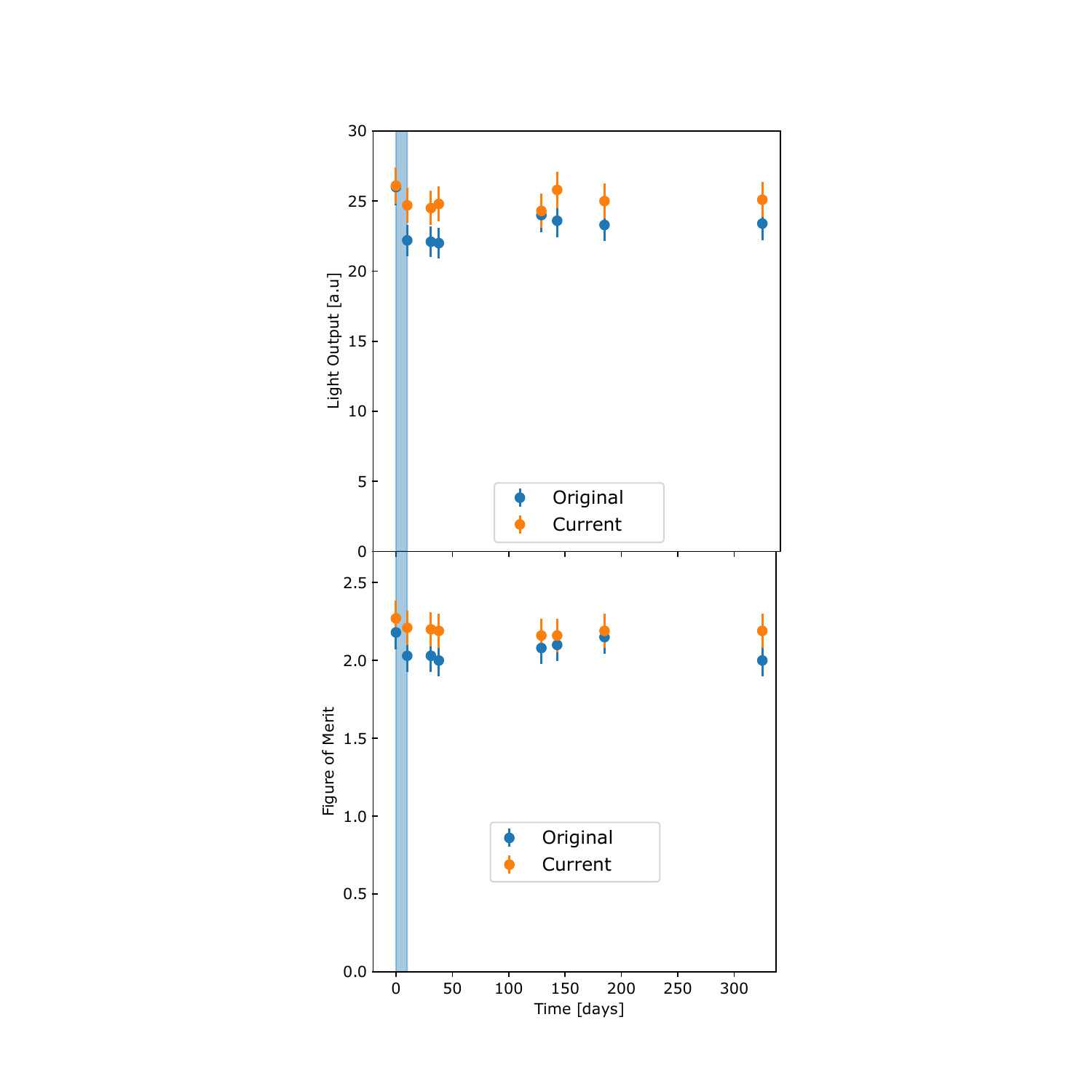}
    \caption{\LO{} and \FoM{} observed for both \EJ{} variants during almost one year of continuous measurements in an air environment. The blue band represents the short period of aging tests through heat treatment at \SI{60}{\degreeCelsius}}
    \label{Fig:aging}
\end{figure}

The two metrics used to asses the stability of the plastics are the relative light output (\LO{}), and the PSD \FoM{}, measured at different points in time and representing different stages of the aging process. \LO{} was determined using the Compton-edge of \Cs{}, and \FoM{} using neutron-gamma data from \Cf{}, both in a similar fashion as in Secs.~\ref{Sec:AttLength} and \ref{Sec:NeutronStudies}. The evolution of \LO{} after the aging tests can be observed in the upper graphs in Fig.~\ref{Fig:aging}, where the ``Original" variant shows a decrease in \LO{} following a temperature treatment (\SI{60}{\degreeCelsius} in an air environment). In contrast, the ``Current" variant experiences a milder deterioration in \LO{}. The \FoM{} values in the lower graphs in Fig.~\ref{Fig:aging} follow a consistent pattern, suggesting that there are no substantial effects on the inherent \PSD{} mechanism. After prolonged exposure to air over a period of 19 weeks, no significant performance effects can be noted. The uncertainties for the measurements shown in Fig.~\ref{Fig:aging} have an associated relative systematic uncertainty of 5\%.

Generally, we can conclude from these tests that, under the relatively harsh environment of the heat treatment in air, the tested variants maintained good \LO{} and \PSD{} performance after 19 weeks (almost a half-year) of observations.

%% file: Conclusions.tex
\section{Conclusions}
In this manuscript the performance of the novel \Li{}-doped \PSD{}-capable EJ-299-50 plastic scintillator developed by  %Lawrence Livermore National Laboratory and 
Eljen Technology has been characterized in the context of  casting appropriate for large detector systems. The results present a promising technology for, e.g., future reactor-antineutrino detectors. Light output and attenuation length for this plastic are similar to liquid counterparts, while solid state permits geometrical versatility and increased mobility. The material has also proven adequate for neutron identification, which could be useful for future neutron multiplicity, directionality, and spectroscopy detectors. Finally, aging tests performed on variants of the main composition show relative stability over time, crucial for the future of long-term uses of plastic scintillators.

\section{Acknowledgements}
This work was performed under the auspices of the U.S. Department of Energy by Lawrence Livermore National Laboratory under Contract DE-AC52-07NA27344. This work was supported by the LLNL-LDRD Program under Project No. 20-SI-005 and by the U.S. Department of Energy Office of Defense Nuclear Nonproliferation Research and Development. LLNL-JRNL-861520.